\documentclass[
 prb,
 amsmath,amssymb,
 aps
]{revtex4-2}
\usepackage{graphicx}
\usepackage{appendix}
\usepackage{bm}
\usepackage{braket}
\makeatletter
\newsavebox{\@brx}
\newcommand{\llangle}[1][]{\savebox{\@brx}{\(\m@th{#1\langle}\)}%
  \mathopen{\copy\@brx\kern-0.5\wd\@brx\usebox{\@brx}}}
\newcommand{\rrangle}[1][]{\savebox{\@brx}{\(\m@th{#1\rangle}\)}%
  \mathclose{\copy\@brx\kern-0.5\wd\@brx\usebox{\@brx}}}
\makeatother
\begin{document}
\title{Ground states of a family of frustrated spin models for quasicrystals and their approximants}

\author{Anuradha Jagannathan}
\affiliation{
Universit\'{e} Paris-Saclay, CNRS, Laboratoire de Physique des Solides, 91405 Orsay, France
}
\date{}

\begin{abstract}
Many new families of quasicrystal-forming magnetic alloys have been synthesized and studied in recent years. For small changes of composition, the alloys can go from quasiperiodic to periodic (approximant crystals) while conserving most of the local atomic environments.  Experiments show that many of the periodic approximants order at low temperatures, with clear signatures of ferromagnetic or antiferromagnetic transitions, and also in some cases undergo non-equilibrium spin glass transitions. In contrast, the quasicrystals are mostly found to be spin glasses. Systematically studying these alloys could help elucidate the role played by quasiperiodicity in (de)stabilizing long range magnetic order. In this work, we study cluster spin models with the aim of understanding the mechanisms behind various types of long range magnetic ordering in approximants and quasicrystals. These models embody key features of real systems, and to some extent are analytically tractable, both for periodic and quasiperiodic cases. For the quasicrystal, we describe two novel magnetic phases with quasiperiodic ordering. Our results should serve to motivate further studies with detailed numerical explorations of this family of models. 
\end{abstract}

\maketitle

\section{Introduction} Many new magnetic quasicrystals with well-defined local moments have been synthesized in the last two decades. Given their noncrystallographic symmetries and absence of translation invariance, one might well expect to find novel forms of magnetic ordering in these systems. In practice, however, ordering into an equilibrium low temperature magnetic phase generally does not occur. The notable exceptions are the ferromagnetic icosahedral AuGaGd \cite{tamura2021} alloy, and the recently reported antiferromagnetic i-AuInEu \cite{tamura2025}. These examples underscore the fact that ordering is rarely seen in quasicrystals. Their behavior stands in contrast to closely related periodic structures called approximant crystals, which often do display long range magnetic order \cite{suzuki,shiino}.  One of the objects of this paper is to understand some of the reasons behind this difference. We examine why magnetic order is difficult in quasicrystals, how long range quasiperiodic structural order affects the ordering of spins, and what new types of equilibrium magnetic phases may be possible in a quasicrystal.

In principle, quasicrystals can be expected to host many forms of magnetic ordering. The simplest magnetic states are commensurate with the lattice, such as the Neel-type collinear state in a Heisenberg antiferromagnetic nearest neighbor model in the Penrose rhombus tiling or the eight-fold Ammann-Beenker tiling. These tilings are built from tiles with an even number of sides, whence their bipartite property and absence of frustration. The situation is less clear for frustrated models. The spin systems that we will be concerned with in this paper are frustrated in the sense that the pairwise spin-spin interactions cannot all be simultaneously satisfied. Frustration in these systems occurs for two reasons. The first is due to the geometry of the structures, due to triangles of interacting spins. The second is due to competition between longer range indirect magnetic interactions due to conduction electrons. These well-known RKKY interactions in periodic metals \cite{rkky}, are expected to exist in quasicrystals both on general grounds, and based on computations done on tiling models \cite{jaga1994,thiem,roche}. 

There have been a number of numerical studies of magnetism in quasicrystals to date. Classical spin models for Ising, planar and Heisenberg spins have been studied in a variety of quasiperiodic lattices \cite{ogueydun,ledue,reid,vedmedenko, matsuo,sugimoto2016,miyazaki,qvarngard}. In one study of particular relevance to our problem, a periodic approximant in which spins interact via long-range RKKY couplings was studied \cite{miyazaki} by Monte Carlo simulation. A variety of magnetic phases were found as the RKKY coupling strengths are varied by tuning the electron density. The phase diagram includes ferromagnetic, antiferromagnetic and a number of incommensurate/disordered phases. The nature of the complex magnetic orderings remain however to be completely determined. Further computational challenges are posed when the model is extended to larger approximants. An alternative simpler approach to modelling experimental systems is therefore desirable. The 2D toy models which we will introduce allow for considering larger periodic systems. They allow for theoretical analyses which can be useful to guide future numerical investigations and moreover give an intuitive picture for ordering in 3D models.

It can be mentioned, for completeness, that quantum spin models have been studied in quasicrystals. Magnetism in quasiperiodic tilings for unfrustrated Heisenberg nearest neighbor models have been studied. Such models describe, for example, a  S=1/2 antiferromagnetic quasicrystal -- where one can ask whether quantum fluctuations could destroy Neel ordering in the ground state. The answer, that Neel order indeed persists, has been given for the spin $\frac{1}{2}$ Heisenberg model in the Penrose and Ammann-Beenker (octagonal) 2D quasiperiodic tilings \cite{wessel,szallas}.  Exploiting the scale invariance of quasicrystals, a renormalization group has been used to compute ground state properties in the Ammann-Beenker tiling \cite{jagaPRL}. Other quantum studies  include:  dimer ground state in a quasiperiodic antiferromagnet \cite{ghosh}, classical and quantum dimer model solutions for 2D tilings \cite{flicker,singh}, and a Kitaev type model on a structure based on the Penrose tiling with a spin-liquid ground state \cite{kim}. This said, while quantum models are of interest for their fundamental aspects, for our present experimental systems, classical spin models such as the ones discussed in this paper should suffice.

This paper is organized as follows: Sec.II introduces a family of cluster spin models. In Sec.III results for the magnetic ground states for different periodic structures will be described, in increasing order of their complexity. Sec.IV discusses a quasiperiodic structure and describes two novel different types of long range ordered magnetic phases. Sec.V concludes with a discussion and perspectives.

\section{Cluster models on square triangle tilings}
The experimental systems that we are seeking to understand are metallic alloys containing rare earth ions. The local moments (spins) are therefore embedded in a conducting medium, which gives rise to long range indirect magnetic interactions between the spins. The spins are primarily located on icosahedral clusters, whose locations can be periodic or quasiperiodic. Our goal is to study simplified models of real systems where clusters of spins interact. The structures may be quasiperiodic or periodic crystals whose local environments resemble some of the ones found in the quasicrystal. For greater theoretical and numerical tractability, we have chosen to consider two dimensional structures. It must be stressed that, although in 2D continuous symmetry cannot be broken at finite temperature -- there is no long range order -- a quasi long-range order can be found at low temperatures and the $T=0$ phase diagrams of 2D models give a general understanding of the different types of magnetic phases. Among 2D quasicrystalline structures, the rich set of approximant and quasiperiodic structures belonging to the family of square triangle tilings are very well suited for our toy models. 

\begin{figure*}[h!]
\includegraphics[width=0.3\textwidth]{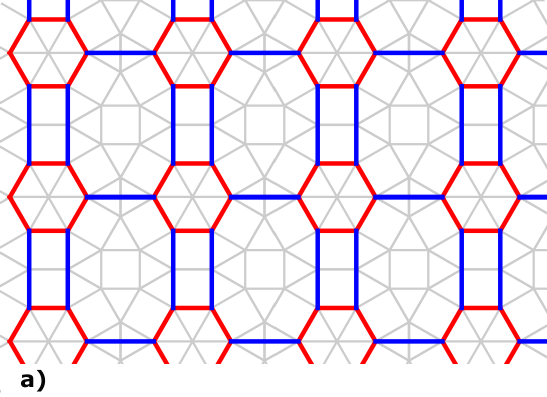} \hskip 0.5cm
\includegraphics[width=0.25\textwidth]{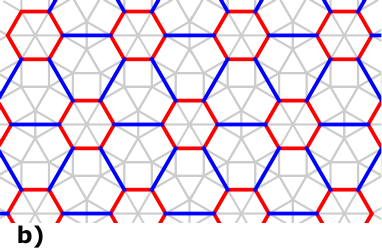} \hskip 0.5cm
\includegraphics[width=0.3\textwidth]{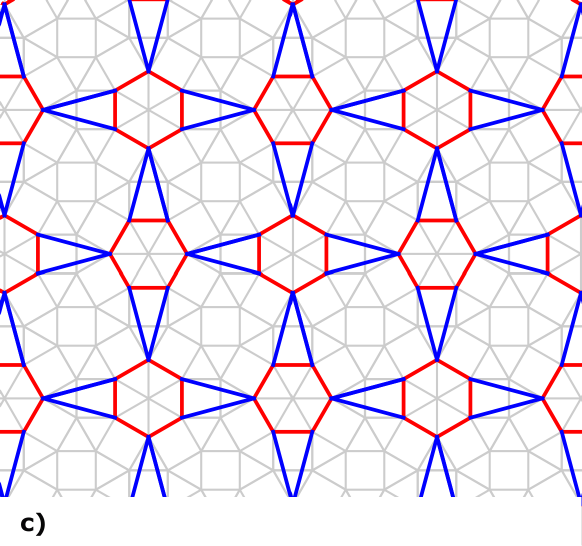} 
\vskip 0.5cm
\includegraphics[width=0.4\textwidth]{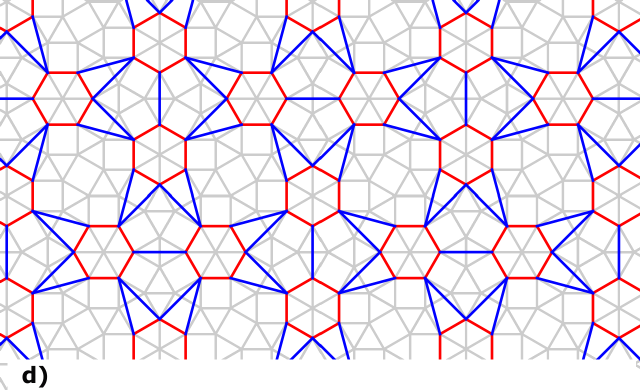}\hskip 0.5cm
\includegraphics[width=0.4\textwidth]{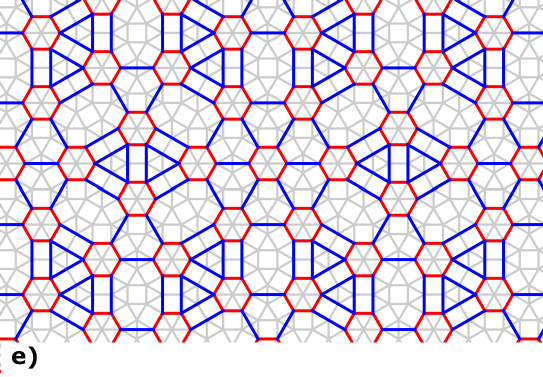}
\vskip 0.5cm
\caption{Five examples of spatial arrangements of hexagonal clusters of 6 spins and inter-cluster bonds. The figures show close-ups of local environments for five models. In each case, intra-cluster bonds are shown in red, inter cluster bonds in blue. The underlying square-triangle lattice is shown in grey. a) Square array of clusters  b) The hex-$\triangle$ model, a hexagonal array of clusters c) The hex-$\square$ model, a square array with staggered orientation of clusters d) The hex-$\sigma$ model, with clusters arranged on a sigma lattice and e) The hex-QC model, with clusters on vertices of a 6-fold quasiperiodic tiling.  }
\label{fig:lattices}
\end{figure*}

The family of square triangle tilings includes periodic structures, from the triangular and square lattices to more complex unit cell structures, deterministic dodecagonal quasicrystals \cite{kawa,stampfli,niizeki,gahler,hermisson}, and random dodecagonal quasicrystals \cite{henley} among others (see review in \cite{marianne}). Dodecagonal tilings have been experimentally observed in solid state alloys, as reviewed in \cite{ishimasa}, and also in soft matter systems \cite{soft1,soft2}. In the present study, we only focus on deterministic square triangle tilings, and in particular those which contain spatially distinct clusters of hexagons composed of six triangles. Two orientations are possible for hexagons --  those made of triangles having one vertical edge ($v$-hexagon), and those made of triangles having one horizontal edge ($h$-hexagon). 
Fig.\ref{fig:lattices} gives examples of cluster models defined on square triangle tilings. In each case, the underlying square triangle tiling is shown in light grey. The six nearest neighbor bonds for each cluster $J_1$ are shown by bold red lines while inter-cluster bonds are drawn in blue. For simplicity, in each of the cases, only the shortest bonds needed to form a connected 2D structure are kept. Note that next nearest neighbor bonds within a given cluster are not considered, these as well as longer range inter-cluster interactions can be added in further refinements of the models.  The first four structures are periodic, while the fifth is quasiperiodic. Many of these structures are novel and have not been studied before in the literature of frustrated spin systems, to our knowledge. They are briefly introduced here as follows:

\begin{itemize}
\renewcommand\labelitemi{}
\item -- In Fig.\ref{fig:lattices}a) the centers of hexagons lie on a square lattice of side $(2+\sqrt{3})a$ where $a$ is the side of the basic tile (in grey). Along the $x$ (horizontal) direction, clusters are connected by a single intercluster bond, while along $y$ (vertical) direction they are linked by a pair of intercluster bonds. The resulting spin model has a rectangular symmetry. The intercluster bonds (in blue) have two different lengths $l_2=\sqrt{3}a$ and $l_3=2a$. 
\item -- Fig.\ref{fig:lattices}b) shows a model where the hexagons form a triangular lattice. We term the corresponding spin model the $H-\triangle$ model. The inter-cluster bonds correspond to a distance $l_2$. 
\item -- In Fig.\ref{fig:lattices}c) the hexagons lie on a square lattice of side $(2+\sqrt{3})^{1/2}a$. This structure which we term the $hex-\square$ model, has hexagons of both $h$ and $v$ orientations.  
\item --In Fig.\ref{fig:lattices}d) the hexagons lie on the vertices of a well-known Archimedean lattice called the sigma lattice in the literature of Frank-Kasper phases. This structure will thus be called the $hex-\sigma$ model. The intercluster bonds shown correspond to distances $l_1$ and $l_3=\sqrt{2+\sqrt{3}}a\approx 1.93a$. 
\item -- Fig.\ref{fig:lattices}e) shows a cluster model based on a quasiperiodic tiling. The underlying tiling, shown in grey, is quasiperiodic (infinite unit cell) with overall 6-fold symmetry. As will be seen in Sec.IV this tiling can be obtained by a substitution method, and possesses a discrete scale invariance with scale factor $\lambda=2+\sqrt{3}$. It contains hexagonal clusters which are all of $h$ type (thus differing from the well-known square-triangle dodecagonal quasicrystal which has both $h$ and $v$ hexagons in equal numbers). 
\end{itemize}

The five examples given in Figs.\ref{fig:lattices} illustrate the types of local environments and the increase of their complexity when going from periodic lattices towards quasiperiodic structures. Some or all of the local bond configurations present in the simpler lattices will be seen to be present in more complex cases. In the present family of models, there are three ways in which two hexagons are coupled -- i) a single $J_2$ bond, or ii)  two $J_2$ bonds or iii) a $J_1-J_2-J_2$ triangle. Similarly, triplets of hexagons can be coupled in a variety of ways. The 6-fold quasicrystal shown has a relatively simple bond configuration as compared to the dodecagonal quasicrystal (not shown), making it possible to find the ground state solutions described in Sec.IV. 

For each of the cluster spin models, we consider a Heisenberg Hamiltonian of the form

\begin{eqnarray}
    H =- J_1\sum_{\langle i,j\rangle} {\bf{S}_i \cdot \bf{S}_j} - J_2\sum_{\llangle i,k\rrangle} {\bf{S}_i \cdot\bf{S}_k}
    \label{eq:heisenberg1}
\end{eqnarray}
where the first sum is over pairs of sites $\langle i,j\rangle$ linked by intra-cluster bonds (red bonds), and the second sum $\llangle i,j\rrangle$ is over inter-cluster bonds (blue bonds). We will assume for simplicity that $J_2$ coupling is the same for all intercluster bonds even if their bond lengths are slightly different (in Fig.\ref{fig:lattices}a), one sees that the intercluster bond lengths along $x$ and $y$ are not exactly equal). One can note that aside from trivial cases, the properties of $H$ are governed by the ratio of couplings, and by their signs. Our goal is to discuss ground states as a function of $J_1$ and $J_2$ for each of the models. The $++$ quadrant or fully ferromagnetic (FM) case is trivial, the ground state is the uniform ferromagnet. We only discuss the three other quadrants, where depending on the frustration, much more complex forms of ordering can occur.

The class of models Eq.\ref{eq:heisenberg1} considers only isotropic Heisenberg interactions between spins. It should be noted however, that in real experimental alloys of interest, the symmetry of spin-spin interactions will depend on the nature of the rare earth ion. The isotropic models above could describe magnetic alloys in which the rare earth ions have no orbital moment, such as Gd$^{3+}$ or Eu$^{2+}$. However, for other cases, such as Tb$^{3+}$ ions in AuGaTb alloy \cite{labib}, crystal field effects giving rise to anisotropic interaction terms must be taken into account. Those anisotropic interactions and the resulting phase diagrams are beyond the scope of the present work. 

\section{Periodic models}
In the first structure, shown in Fig.\ref{fig:lattices}a), clusters are coupled by a single $J_2$ bond along the $x$ direction and double $J_2$ bonds along the $y$ direction. It is not frustrated for any choice of sign of $J_1$ and $J_2$. Its structure made of even-sided loops ensures that all bonds can be satisfied, whatever the sign of the couplings. Ground states are collinear states that can be obtained from the uniform FM state by flipping all spins of a given sublattice. The structure shown in Fig.\ref{fig:lattices}a) does not require any further discussion. We now turn to the remaining structures, which are  frustrated in one or more of quadrants of the phase diagram.

\subsubsection{The $hex-\triangle$ model}
The structure shown in Fig.\ref{fig:lattices}b), called the $hex-\triangle$ model, has $h$ hexagons placed in a triangular array. It can be seen that the coordination number of the sites is 3, that is, the same as that of the honeycomb lattice. The Hamiltonian of Eq.\ref{eq:heisenberg1} has a sublattice symmetry, as follows: the sites of each hexagon are assigned to sublattices $\mathcal{S}_1$ and $\mathcal{S}_2$ in alternating fashion, using the same convention for all hexagons. The Hamiltonian only couples sites on different sublattices. This chiral symmetry present in the Hamiltonian of Eq.\ref{eq:heisenberg1} gives the invariance under $J_{ij}\rightarrow -J_{ij}$ and changing the signs of the spins on one sublattice, that is, $\bf{S}_j \rightarrow -\bf{S}_j$ for $j\in \mathcal{S}_2$ while keeping spins on sublattice 1 fixed. 

The ground state for the $--$ quadrant (antiferromagnetic couplings) is simply obtained from the ferromagnetic state, by flipping all spins on one sublattice. In this Néel type antiferromagnetic state, each of the hexagonal clusters has an alternating collinear ordering of the spins. The magnetic and the structural unit cells are the same. All the bonds are satisfied, the system is not frustrated in the two quadrants $++$ and $--$, and the ground state energy per spin is (for $J_1\neq 0$)
\begin{eqnarray}
E_{0}^{hex-\triangle}(r;++) = E_{0}^{hex-\triangle}(r;--) = -NJ_1 S^2 (1+\frac{1}{2}r)
\end{eqnarray}
where $N$ is the number of spins. The parameter $r$ here and throughout the paper is defined by $r=\vert \frac{J_1}{J_2}\vert$ (for $J_1\neq 0$).

Frustration, however, does arise when the couplings are of opposite signs. The ground states in the two quadrants of mixed sign $-+$ and $+-$ are again related by symmetry so we need only consider one of these sectors. Let us consider $+-$ (ferromagnetic $J_1$ antiferromagnetic $J_2$). The ground states are deduced by the following continuity argument. In the limit of very large $J_1$ (that is, $r \ll 1$), the spins in each hexagon minimize the energy by aligning in parallel. The hexagons are coupled by $J_2$ bonds in a triangular array. The result is a state that is closely related to the well-known 3-color ground states of the antiferromagnetic Heisenberg model on the triangular lattice. The three colors (say red, blue and green) correspond to three spin directions oriented at angles of $2\pi/3$ with respect to each other, with ${\bf{S}}_{r}+{\bf{S}}_{b}+{\bf{S}}_{g}=0$. 

\begin{figure*}[h!]
\includegraphics[width=0.65\textwidth]{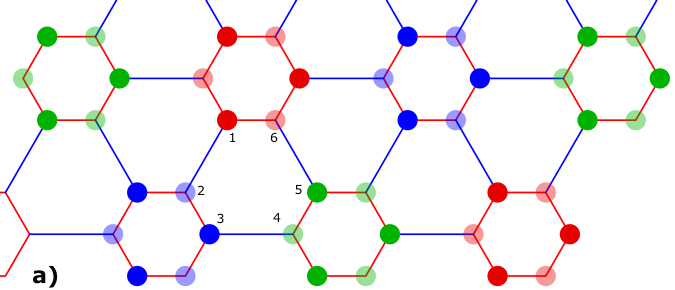} \hskip 1cm
\includegraphics[width=0.2\textwidth]{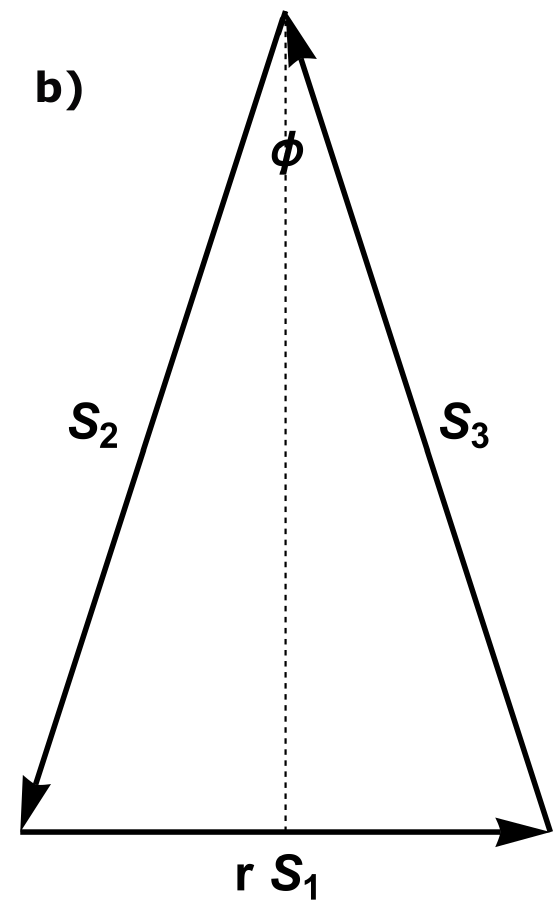}
\caption{a) A ground state for the $hex-\triangle$ model in the $+-$ sector showing spin orientations. The orientations for spins on sublattice $\mathcal{S}_1$ are indicated by circles of darker shades, colored red (spins aligned with an axis of reference), blue (at an angle of $2\pi/3$) and green (at an angle of $4\pi/3$). The orientations for spins on sublattice $\mathcal{S}_2$ are indicated by the circles of lighter shades, directed along $-\theta$ (red), $2\pi/3-\theta$ (blue) and $4\pi/3-\theta$ (green). b) Geometrical representation of constraint satisfied by spins in the ground state of $H_{tri}$ defined in Eq.\ref{eq:separateham}, for $r\le 2$.}
\label{fig:hextristate}
\end{figure*}

As $r$ is increased, this state evolves so as to have a 3 color state on each of the sublattices (and thus six colors in total). This occurs as follows: in each hexagon, all the 3 spins of sublattice $\mathcal{S}_1$  have the same color (red, blue or green), defining thereby the color of each hexagon. The colored hexagons map to the 3-color ground states of the triangular lattice. Next, we determine the directions of the remaining spins on sublattice $\mathcal{S}_2$, which depends on the value of $J_2$. To do this, one considers a loop of six spins belonging to three hexagons -- those numbered 1 through 6 in Fig.\ref{fig:hextristate}a). Assuming that sublattice 2 spins are rotated with respect to their neighboring spins by an angle $\theta$, the energy of the loop of 6 spins is $-3J_1 \cos\theta + \frac{3}{2}\vert J_2 \vert \cos(2\pi/3+\theta)$. 
Minimizing the energy, one gets the angle of rotation as a function of $r$,
\begin{eqnarray}
\theta &= &\tan^{-1}\left(\frac{\sqrt{3}r}{4+r}\right)
\end{eqnarray}
Note that, as one would expect, the angle $\theta$  tends to zero in the limit of $r \rightarrow 0$ (when the $J_1$ ferromagnetic bonds are dominant), and tends to $\pi/3$ when $r\rightarrow \infty$ (when the intercluster antiferromagnetic bonds are dominant). Extending this solution to the entire lattice, the ground state $E_0$ in the $+-$ quadrant is obtained to be
\begin{eqnarray}
E_{0}^{hex-\triangle}(r;+-) &=& - \frac{1}{2}N J_1 S^2 \sqrt{4+r(2+r)} 
   \label{eq:gs1}
\end{eqnarray}
This state is coplanar at $T=0$ for the same reason as the classical Kagome antiferromagnet, due to the larger number of zero modes for coplanar configurations as compared to noncoplanar ones \cite{chalker}. That is, all the spin vectors lie in the same plane, which can of course have an arbitrary angle with respect to the plane of the lattice itself. The spins on the A-sublattice are aligned along one of the three principal directions, at angles of $2n\pi/3$ ($n=0,1,2$). Such a state is illustrated in Fig.\ref{fig:hextristate}a) using three colors for the three directions. The B-sublattice spins are additionally rotated by an angle $-\theta$ with respect to their neighbors, and are indicated by lighter shades.   
The magnetic unit cell of such a 6-color ground state is $\sqrt{3} \times \sqrt{3}$ bigger than the structural unit cell. When couplings have the opposite signs ($-+$ quadrant) the ground state configurations can be deduced from the above using the symmetry property, and the ground state is unchanged, that is, $E_{0}^{hex-\triangle}(r;-+) = E_{0}^{hex-\triangle}(r;+-)$ is given by Eq.\ref{eq:gs1}.

To resume, for the $hex-\triangle$ model, the magnetic states in the $J_1-J_2$ phase diagram can be collinear or coplanar but are all commensurate with the lattice. This is not always the case for the following structures, as discussed in the following subsections. 

\subsubsection{The $hex-\square$ model}
The lattice shown in Fig.\ref{fig:lattices}c) is composed of alternating hexagons of $h$ and $v$ type. One sees that it has geometrical frustration due to triangles composed of two $J_2$ bonds and one $J_1$ bond. We define two sublattices as follows: all sites of $h$-hexagons belong to sublattice $\mathcal{S}_1$, while those on $v$-hexagons belong to sublattice $\mathcal{S}_2$. Then the Hamiltonian has the following symmetry: for $J_1$ fixed, the ground state energy is invariant under $J_2 \rightarrow -J_2$ along with $\vec{S}_i \rightarrow -\vec{S}_i$ for all the spins belonging to one of the sublattices. 

\begin{figure*}[h!]
\includegraphics[width=0.25\textwidth]{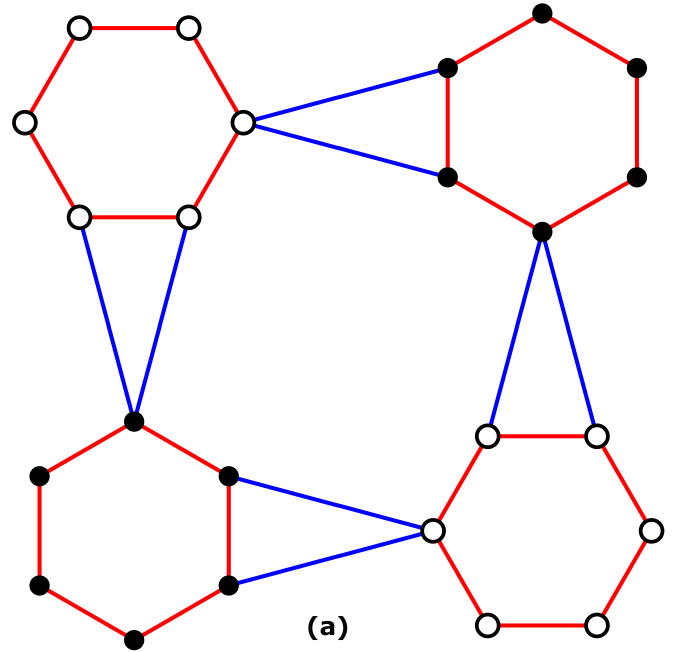} \hskip 1cm
\includegraphics[width=0.25\textwidth]{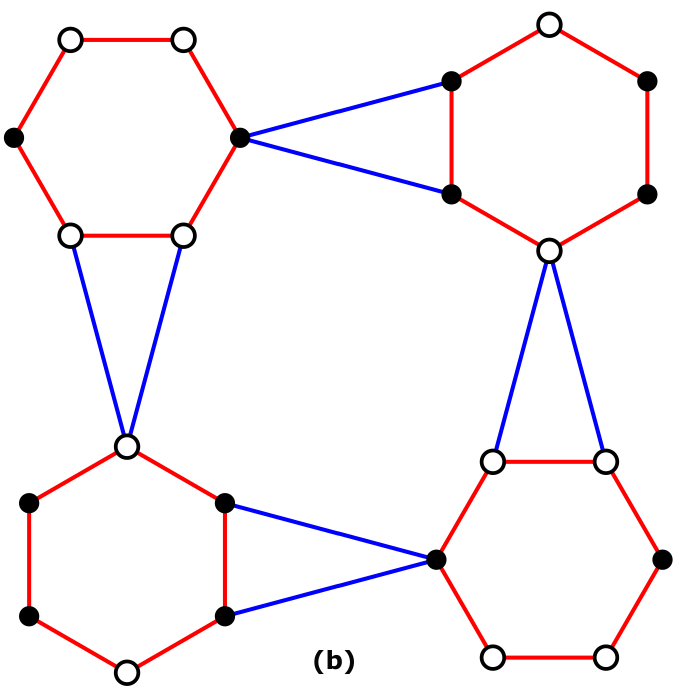}  \hskip 1cm
\includegraphics[width=0.25\textwidth]{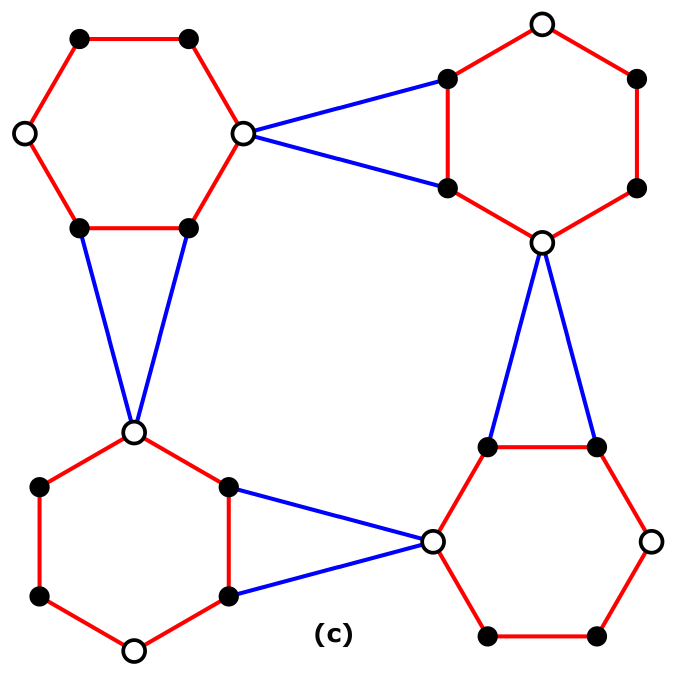} 
\caption{Three out of the four collinear states of the $hex-\square$ lattice (a trivial ferromagnetic phase is not shown). a) The $+-$ quadrant cluster-antiferromagnet phase c-AFM1, b) the $-+$ quadrant cluster-antiferromagnet phase c-AFM2 and c) the $--$ quadrant cluster-ferromagnet phase c-FM2.}
\label{fig:collinear}
\end{figure*}

We start with the two unfrustrated sectors. For ferromagnetic couplings $++$, the ground state is of course the uniform FM phase. By symmetry, the ground state for the $+-$ sector is had by flipping spins on all the $v$-hexagons. The result is the collinear cluster antiferromagnet c-AFM1 shown in Fig.\ref{fig:collinear}a). The ground state energies are  
\begin{eqnarray}
  E_{0}^{hex-\square}(r;++)=E_{0}^{hex-\square}(r;+-) = -NJ_1 S^2(1 + \frac{2}{3}r)
    \label{eq:gs2}
\end{eqnarray}


The model is frustrated for antiferromagnetic $J_1$, whatever the sign of $J_2$. However, the ground state energy and manifold of states can still be calculated exactly because the Hamiltonian can be separated into two parts. The first is a sum over single bonds which do not belong to any triangle, and the second is a sum over all triangles consisting of three bonds, as follows
\begin{eqnarray}
    H^{hex-\square} &=&  \sum H_{b} + \sum H_{tri}  \nonumber \\
H_{b} &=& J_1 {\bf{S}}_{1} \cdot {\bf{S}}_{2}  \nonumber \\
 H_{tri} &=&   J_1 {\bf{S}}_{2} \cdot {\bf{S}}_{3} + J_2 ( {\bf{S}}_{1} \cdot {\bf{S}}_{2}  + {\bf{S}}_{1} \cdot {\bf{S}}_{3})  
   \label{eq:separateham}
\end{eqnarray}
It can be seen above, and checked in Fig.\ref{fig:lattices}c), that every triangle has one $J_1$ and two $J_2$ bonds. To determine ground states of $H^{hex-\square}$, we start by finding the lowest energy states for a single triangle. We assume below that $J_2<0$, but solutions for the opposite sign can be easily determined by symmetry. For any given triangle, the energy terms can be rewritten as follows    
\begin{eqnarray}
    H_{tri} &=&  \frac{J_1}{2} (r {\bf{S}_1 + \bf{S}_2+ \bf{S}_3})^2 - J_1\left(1 + \frac{r^2}{2}\right)S^2
   \label{eq:triangle}
\end{eqnarray}
The lowest energy of the system is obtained by minimizing the term $(r {\bf{S}_1 + \bf{S}_2+ \bf{S}_3})^2$. For $r \leq 2$, the spins achieve this by satisfying the condition 
\begin{eqnarray}
r{\bf{S}}_1 + {\bf{S}}_2+ {\bf{S}}_3 =0   
\label{eq:constraint}
\end{eqnarray}
Representing the sum of vectors graphically as in Fig.\ref{fig:hextristate}b), one sees that $\bf{S}_2$ and $\bf{S}_3$ must be oriented at angles $\pm(\frac{\pi}{2} +\phi/2)$, with respect to $\bf{S}_1$, where $\phi = 2\arcsin(r/2)$. The three spins lie in a plane, arbitrary. Two minimal energy solutions of opposite chiralities are possible for a given triangle. To find the minimum energy state of the whole lattice, these solutions must be matched from one unit cell to the next, while also minimizing the energy of all of the single bonds. The result, for $r \leq 2$, is a coplanar state with spiral ordering along one of the diagonals. There are two helicities for each direction, and thus these states are 4-fold degenerate. 
The spiral ordering in the lattice is related to the spiral states of the classical limit of the well-known Shastry-Sutherland model \cite{ssm}, to which the present model can be mapped. 

\begin{figure*}[h]
\includegraphics[width=0.45\textwidth]{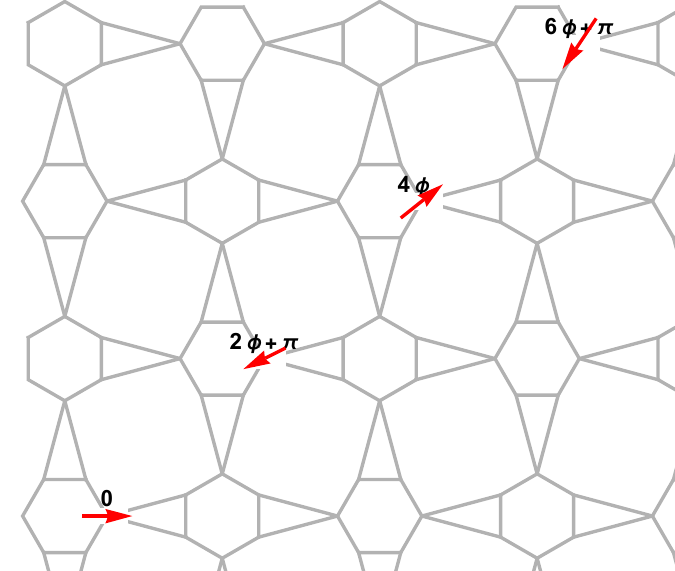} \hskip 1cm
\includegraphics[width=0.45\textwidth]{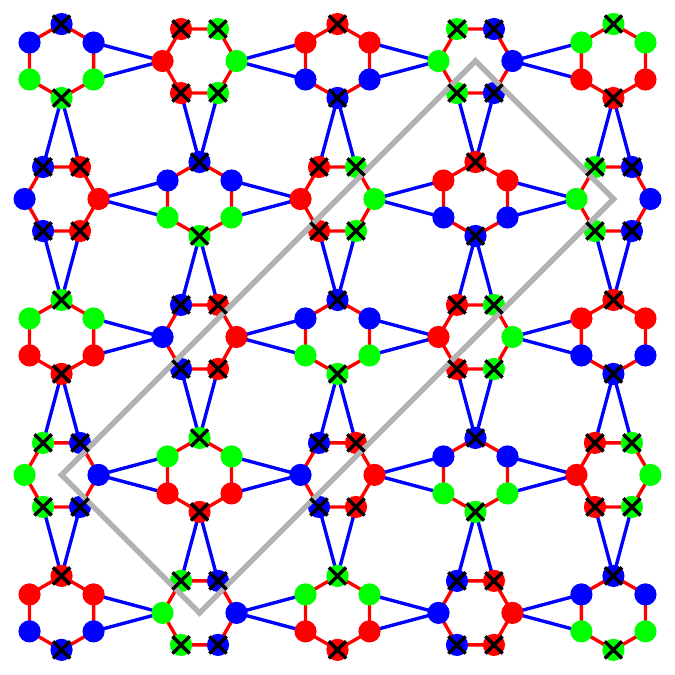}\hskip 1cm
\caption{ (Left) The spiral state is illustrated here by showing the variation of angle of orientation for one of the spins of the unit cell along the diagonal direction. The state is periodic along the direction perpendicular to the spiral. The angle $\phi$ is defined in the text. (Right) Spin configuration in the ground state for $J_1=J_2$, when $\phi=\pi/6$. In this case the magnetic ground state is commensurate with the lattice. The colors red, green and blue correspond to the spin orientations such that ${\bf{S}}_{r}+{\bf{S}}_{b}+{\bf{S}}_{g}=0$. A black cross on a red circle indicates that the spin points in the opposite direction $-{\bf{S}}_{r}$ and similarly for the other colors.  }
\label{fig:spinconfigs}
\end{figure*}

The spiral is commensurate with the lattice with a period $M$, for $\phi=\pi/M$ ($M$ even).  For the case $J_1=J_2$ in particular, $\phi=\pi/6$. The three spins of each triangle are oriented at angles of 120$^\circ$, giving rise to a 3-color ground state. The spin configuration is illustrated for this commensurate case in Fig.\ref{fig:spinconfigs}b). The colors red, green and blue correspond to the spin orientations such that ${\bf{S}}_{r}+{\bf{S}}_{b}+{\bf{S}}_{g}=0$. A black cross on a red circle indicates that the spin points in the opposite direction $-{\bf{S}}_{r}$ and similarly for the other colors.


For $r\geq 2$, the ground state of $H_{tri}$ is a collinear state with $  {\bf{S}}_2={\bf{S}}_3 = -{\bf{S}}_1$. This solution extended over the entire lattice results in the cluster ferromagnetic c-FM2 state illustrated in Fig.\ref{fig:collinear}c). One can note here a seemingly counterintuitive result -- of a large enough antiferromagnetic intercluster coupling $J_2$ leads in this model to a ferromagnetic phase !

The energies of the spiral and the collinear states are
   \begin{eqnarray}
   E_{0}^{hex-\square}(r;--)  &=& - N\vert J_1\vert S^2 \left(1 + \frac{r^2}{6}\right) \qquad \qquad r\leq 2 \nonumber \\
    &=& - N\vert J_1\vert S^2 \left(\frac{1}{3} + \frac{2r}{3} \right) \qquad \quad  r > 2
    \label{eq:gs3}
\end{eqnarray}
The solutions for the remaining frustrated $-+$ sector are obtained by symmetry. The collinear state for $r>2$ in the $-+$ sector is a cluster antiferromagnetic state where each cluster has four spins up and two spins down, as illustrated in Fig.\ref{fig:collinear}b). The energies are given by Eqs.\ref{eq:gs3} since $E_{0}^{hex-\square}(r;-+)=E_{0}^{hex-\square}(r;--)$.

\begin{figure*}[h] 
\includegraphics[width=0.5\textwidth]{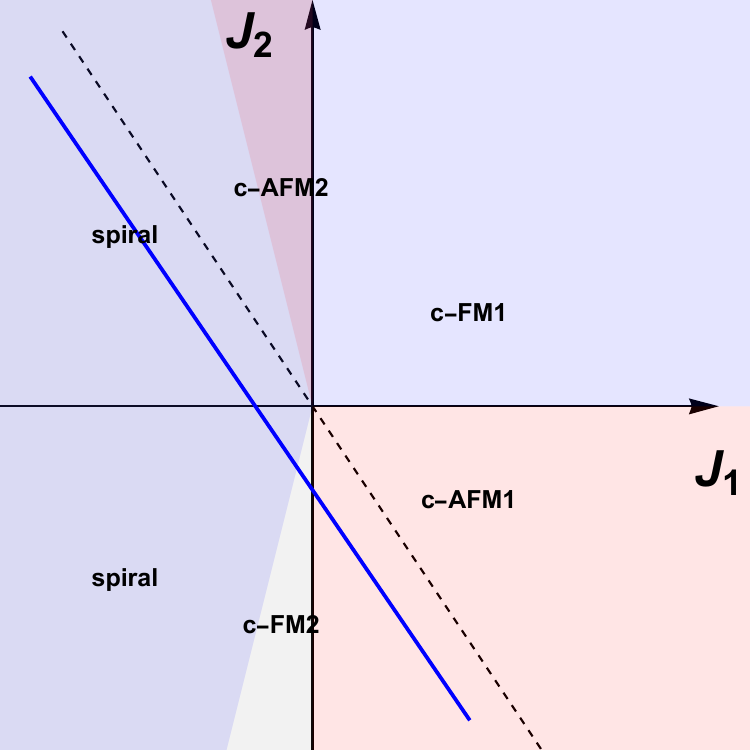}
\caption{ Phase diagram of the $hex-\square$ lattice in the $J_1-J_2$ plane. Ferromagnetic phases are shown in blue and gray-blue, antiferromagnetic phases are shown in red and red-gray. The structures of these phases, which are collinear, are shown in Figs.\ref{fig:collinear}. Spiral phases can be incommensurate, or incommensurate but always periodic along one of the diagonals of the lattice. The dashed line represents the line $T_{cw}=0$. The blue line is a path of fixed negative $T_{cw}<0$ in the $J_1-J_2$ plane, and it is seen to intersect four different phases, depending on the values of the couplings.}
\label{fig:phasediag1}
\end{figure*}

The phase diagram for the model in the $J_1-J_2$ plane is shown in Fig.\ref{fig:phasediag1}. The blue and blue-gray regions are the collinear ferromagnets, where c-FM1 has all spins  parallel with a net spin of 6S per cluster, while c-FM2 has a reduced net spin of 2S per cluster. The red and red-gray regions are the collinear cluster antiferromagnets. The phase c-AFM1 is shown in Fig.\ref{fig:collinear}b), where each cluster has the maximal spin of 6S and the second type of antiferromagnet is c-AFM2, where each cluster has a total spin of 2S. Thus, the staggered moment $M_s$, in c-AFM2 is $\frac{1}{3}$ of that in the c-AFM1 phase. The spiral states in the frustrated sector can be, as already pointed out, commensurate or incommensurate, depending on the value of $r$. 

\subsubsection{The $hex-\sigma$ model}
\begin{figure*}[h]  
\includegraphics[width=0.4\textwidth]{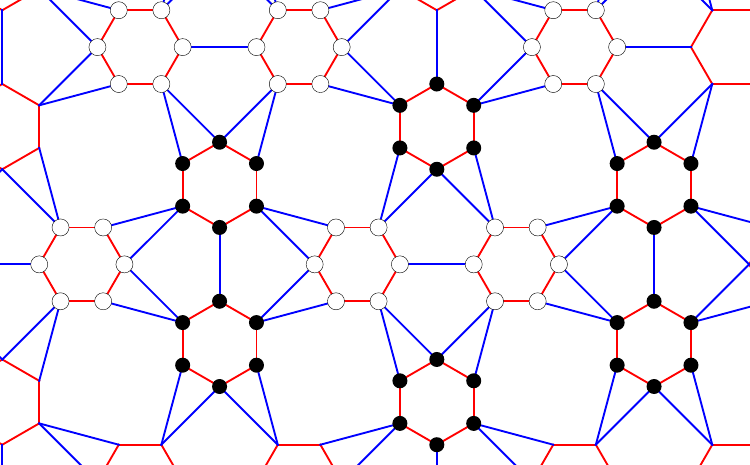}
\caption{Spin configurations in the collinear state in the $+-$ sector of the $hex-\sigma$ lattice. The colors correspond to the up (white) or down (black) spin orientations. This interesting antiferromagnetic state has alternating vertical and horizontal ferromagnetic dimers orientations, as one goes along the $x$ or the $y$ directions. }
\label{fig:hsigma1}
\end{figure*}

The periodic system shown in Fig.\ref{fig:lattices}d), is the $hex-\sigma$ model. It derives from a periodic square triangle lattice with a square unit cell of period $L=(5+3\sqrt{3})/\sqrt{2}$. There are four hexagons (two $h$ and two $v$) per unit cell, whose centers lie on an Archimedean lattice called the sigma lattice. The problem has a sublattice symmetry as follows. We let spins on all the $h$ hexagons belong to sublattice $\mathcal{S}_1$ and the spins on $v$ hexagons belong to sublattice $\mathcal{S}_2$. Then the Hamiltonian is invariant under a change of sign of of $J_2$ accompanied by a flip of spins on one of the sublattices. For the ground states this implies that only two sectors require analysis, since 
\begin{eqnarray}
  E_{0}^{hex-\sigma}(r;++)  &=&  E_{0}^{hex-\sigma}(r;+-) \nonumber \\
  E_{0}^{hex-\sigma}(r;--)  &=&  E_{0}^{hex-\sigma}(r;-+)
\end{eqnarray}
For the unfrustrated sectors, the energy can be written down exactly.
The $+-$ quadrant has a collinear ground state which is related to the trivial ferromagnetic phase by flipping all spins on one of the sublattices, as shown in Fig.\ref{fig:hsigma1}. The figure shows spin orientations for one magnetic period along the $x$ axis, using black shading for clusters of spins pointing up and light grey for spins pointing down for a limiting case of vanishingly small $r$. This ordering can be continued in the $xy$ plane in such a way that there are alternating horizontal and vertical dimers along both the $x$ and $y$ axes. This spin configuration has an energy of
\begin{eqnarray}
    E_0^{(hex-\sigma)}(r;+-) =  -NJ_1 S^2 (1 + \frac{3}{4}r) 
    \label{eq:gs4}
\end{eqnarray}
\begin{figure*}[h] 
\includegraphics[width=0.9\textwidth]{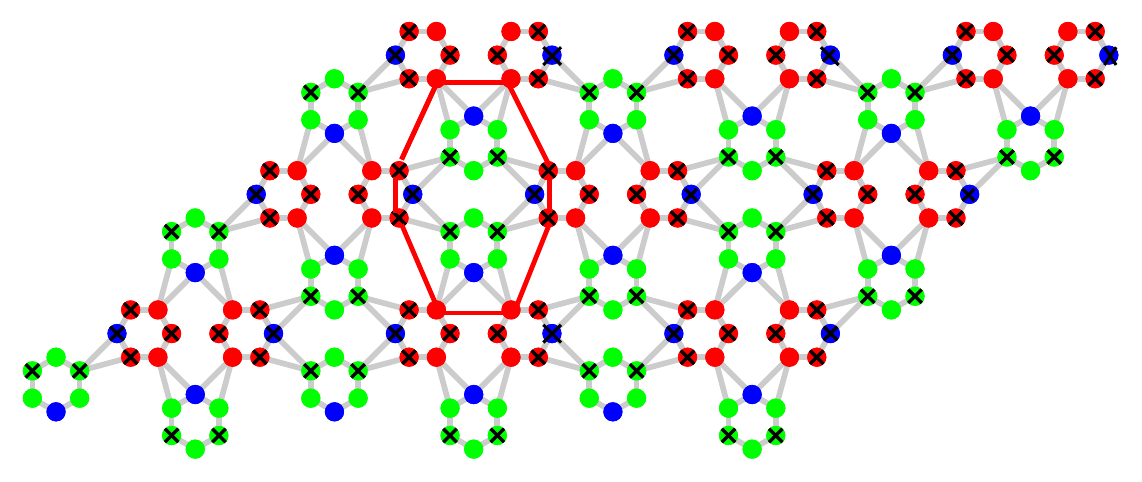} \vskip 1cm 
\caption{One of the ground state spin configurations in the simplified $hex-\sigma$ model (as solved in \cite{jj2025}) showing a periodic configuration. Colored circles red (R), blue (B), green (G) represent three spin directions oriented at $2\pi/3$ with respect to each other. Crosses indicates a change of sign. The red hexagonal shape indicates an area within which all spins can be rotated together around a common axis -- the weathervane mode -- without change of energy. }
\label{fig:hsigma}
\end{figure*}

The situation is more complex in the frustrated sectors. Let us consider the $--$ sector. This model was studied previously for a simplified case \cite{jj2025}, where only the $J_2$ bonds belonging to triangles were kept, while lone $J_2$ bonds are removed. For the resulting ``bond-diluted" model, an exact expression for ground state energy can be written down. Fig.\ref{fig:hsigma} shows one of the ground states for the diluted model when $J_1=J_2<0$. The three colors indicate the direction of the spin, which can be oriented at angles of $0,\pm 2\pi/3$ with respect to some reference direction. A cross on a dot of a particular color indicates the direction opposite to that sublattice direction. Ground states are non-coplanar (have out-of-plane ordering), and are highly degenerate with a macroscopic number of localized zero energy modes (called weathervane modes \cite{weathervane}). This type of soft mode is expected, however, to disappear when the missing $J_2$ bonds are restored. The complete phase diagram for the $hex-\sigma$ model remains to be determined in future investigations.

\subsubsection{Discussion of results for the periodic models.}
The preceding discussion of spin models shows that a rather large variety of magnetic structures can be expected even in the relatively simple cluster models we have introduced. The $hex-\square$ model is a useful case study since it has a rich phase diagram. One can find points in common with 3D models. The cluster antiferromagnets and ferromagnets have their 3D analogs, and without doubt generalized spiral phases must exist in 3D as well. It is thus of interest to consider the implications of our study concerning the experimental results for magnetic approximants. 

Experimental work has shown an interesting systematic behavior in periodic approximants of quasicrystals -- namely, that the nature of magnetic ordering seen correlates with the Curie-Weiss temperature of the alloy. As an example, we consider the results of studies conducted in a family of ternary alloys, Au$_x$Al$_{86-x}$Gd$_{14}$, where $x$ can be varied within a wide range of values \cite{ternarytamura}. Increasing $x$ increases the Fermi wave vector $k_F$, which in turn controls the RKKY interactions between the spins and ultimately, the Curie-Weiss temperature. Appendix A summarizes the different $T_{CW}$ and the resulting magnetic phases as the composition are varied. These data led to the conclusion that small positive values of $T_{cw}$ led to antiferromagnetism, larger positive values to ferromagnetism, and large negative values to the spin glass phase. This tendency has been also seen many other isostructural magnetic approximants \cite{suzuki}, and also in larger unit cell crystalline approximants \cite{tamura2appxt}. 

These experimental observations of the observed correlation between the nature of magnetic orderings and $T_{cw}$ are one of the principal motivations for this theoretical study. One can ask if the phase diagram we obtained for the $hex-\square$ model shows any such correlation with the values of $T_{cw}$. This quantity, proportional to the total coupling of one spin to the others, is $T_{cw} \propto 2J_1+\frac{4}{3}J_2$ for our model. The dashed grey line having a slope of $J_2/J_1=-3/4$ in the figure corresponds to $T_{cw}=0$. The blue line corresponds to a fixed negative value of $T_{cw}$. As can be seen in Fig. \ref{fig:phasediag1}, it passes through different types of ground states. This means that, by varying $J_1$ and $J_2$ while keeping $T_{cw}$ fixed, four different types of low temperature magnetic order can be found -- a spiral phase or two different types of ferromagnet and an antiferromagnetic phase. Similarly, for small enough positive $T_{cw}$, one can expect to find three different kinds of phases, depending on the coupling values. This example underscores the fact that there is no simple general relation between $T_{cw}$ and the type of low temperature magnetic ordering. The systematic behavior seen in experiments presumably arises because of fact that the couplings $J_1$ and $J_2$ are both stem from the RKKY interaction, which places constraints on their values, leading to a smoothly varying $T_{cw}$ as a function of composition (as for simpler crystals in \cite{sakurai}). 

To conclude this section on periodic models, we have seen that frustration plays different roles in the phase diagram, resulting in a variety of collinear and co-planar phases. The $hex-\square$ model has a large region corresponding to spiral order, with both commensurate and incommensurate periods. For $J_1=J_2$, each of the lattices has a different version of a 3-color state. Before moving to the quasicrystal in the next section, we note that all of the incommensurate 1D spiral states of the $hex-\square$ model are periodic in the transverse direction -- that is, spirals form identical parallel strips. Such states cannot be realized in the quasiperiodic model. The ground states found in a large region -- corresponding to practically half the phase space -- of the phase diagram are incompatible with quasiperiodicity.

\section{The 6-fold quasicrystal.}
In this section, we present two novel magnetic states for the quasiperiodic spin cluster model shown in Fig.\ref{fig:lattices}e). The underlying square triangle tiling is indicated by light grey edges. This 6-fold tiling can be generated iteratively starting from a dodecagonal seed, using the method of inflation, as described in Appendix B. An inflation operation transforms the single square(triangle) tile into a $\lambda=2+\sqrt{3}$ times larger patch of square (triangle) shape. It can be seen in Fig.\ref{fig:lattices}e) or in the bigger patch shown in Figs.\ref{fig:threecolorQC}, that the 6-fold quasicrystal possesses hexagons of only one orientation. This stands in contrast with dodecagonal quasicrystals, which have both types of hexagons, consistent with overall 12-fold symmetry (for descriptions of these, see \cite{kawa,stampfli,niizeki,gahler,hermisson}). 

We let $N^{(n)}$ and $V^{(n)}$ denote the number of tiles and number of vertices respectively of the finite patch of the $n$th generation. The number of squares at the $n$th generation of the tiling is denoted $N_s^{(n)}$, and the number of triangles $N_t^{(n)}$. The triangles can appear in two orientations --  $T_1$ triangles (those with a horizontal edge) and $T_2$ triangles (those with a vertical edge) which occur in equal number on the tiling. The 6-fold quasicrystal is obtained in the limit of $n \rightarrow \infty$ iterations of the original dodecagonal seed. Appendix B gives some useful relations for the growth of the patches, and shows in particular that the ratio of number of squares to number of triangles in the infinite tiling is $\sqrt{3}:4$. Note that the number of squares divided by number of triangles is an irrational number, showing that the structure cannot be periodic. 

We assume that the Hamiltonian for the $hex-QC$ model has the standard form of Eq.1, where intracluster bonds $J_1$ and intercluster bonds $J_2$ are shown in red and blue respectively in Fig.\ref{fig:lattices}e).  The number of spins in the $n$th generation tiling is $V^{(n)}=6V^{(n-1)}$.
It can be seen that neighboring hexagons in this structure are coupled in two different ways -- they can be linked by a single $J_2$ bond or linked by two parallel $J_2$ bonds. The two cases correspond to different orientations. The single bonds lie along directions $n\pi/3$ with respect to the $x$ axis, while the double bonds are oriented at $n\pi/3 +\pi/6$ ($n=1,..,6$). The clusters form squares and triangles of $T_1$ and $T_2$ types. This model is non-bipartite and, thus, frustrated in all but the $++$ quadrant.

While a full description of the phase diagram is outside the scope of this paper, we describe two novel kinds of ordered phases in this section. Both these states are commensurate with the underlying structure and predicted to exist in certain regions of the $J_1-J_2$ plane. For other generic values of $J_2/J_1$, however, any magnetic ordering is likely to be fragile or absent. In the absence of periodicity, for example, it would be difficult to stabilize incommensurate spiral states of the type found for the $hex-\square$ model. As a result, ground states with long range order are less likely to occur in the quasicrystal, as compared to the periodic approximants.

\subsubsection{Three-color quasiperiodic ground state.}
Let us consider the $+-$ sector corresponding to ferromagnetic $J_1$ and antiferromagnetic $J_2$. We will describe a long range quasiperiodic 3-color cluster antiferromagnet. This state approaches the exact ground state as $r$ becomes vanishingly small (and non-zero). In the limit of $r \ll 1$, one can define a total spin for each hexagon, ${\bf{T}} =\sum_i {\bf{S}}_i$ where the sum is over the six sites of the hexagon. For small $r$, the spins are aligned along a common direction, so that the total spin is a vector of length $\sim 6S$. I To leading order, for small $r$, one can write an effective Hamiltonian in terms of the total spin variables $\{{\bf{T}}_\mu \}$ to leading order in $r$ as follows :
\begin{eqnarray}
    H^{hex-QC}_{eff} &\approx& - NJ_1S^2 + \sum H_b + \sum H_{tri}   
     \\ \nonumber
     H_{b} &=&  J_2 ~{\bf{S}}_{i} \cdot {\bf{S}}_{j} \\ \nonumber
 H_{tri} &=& J'_{\mu} ~( {\bf{T}}_{1\mu} \cdot {\bf{T}}_{2\mu}  + {\bf{T}}_{2\mu} \cdot {\bf{T}}_{3\mu} +  {\bf{T}}_{1\mu} \cdot {\bf{T}}_{3\mu})  
   \label{eq:hamqc2}
\end{eqnarray}
where the first term is the contribution (constant) arising due to the ferromagnetic intracluster bonds.  The second term, $H_b$ is a sum over all those bonds which do not belong to any triangle. The last term is a sum over all triangular plaquettes, indexed by $\mu$, of the $T$ variables, $\mu=1,...,N^{(n-1)}$. The $H_b$ term involves only a small fraction of sites, compared to the $H_{tri}$ terms, as shown in Appendix C. It can be treated as a perturbation. The three bonds $J'_{\mu}$ of any given triangle are all the same. They can take two different values $\sim J_2/36$ (single bond coupling) or $\sim J_2/18$ (double bond coupling). The ground states for such triangles as we have already seen are the 3-color red-blue-green states. 

\begin{figure*}[h!]
\includegraphics[width=0.45\textwidth]{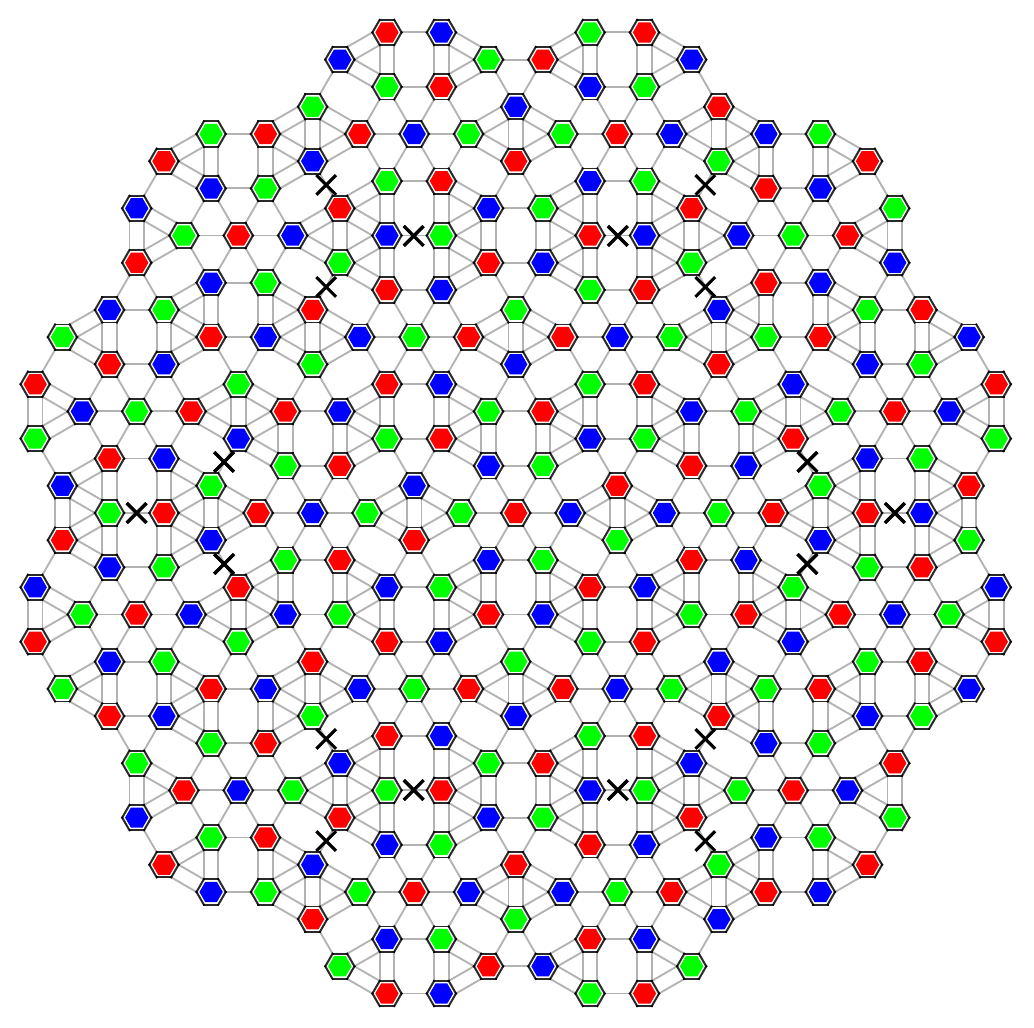} \hskip 1cm
\includegraphics[width=0.45\textwidth]{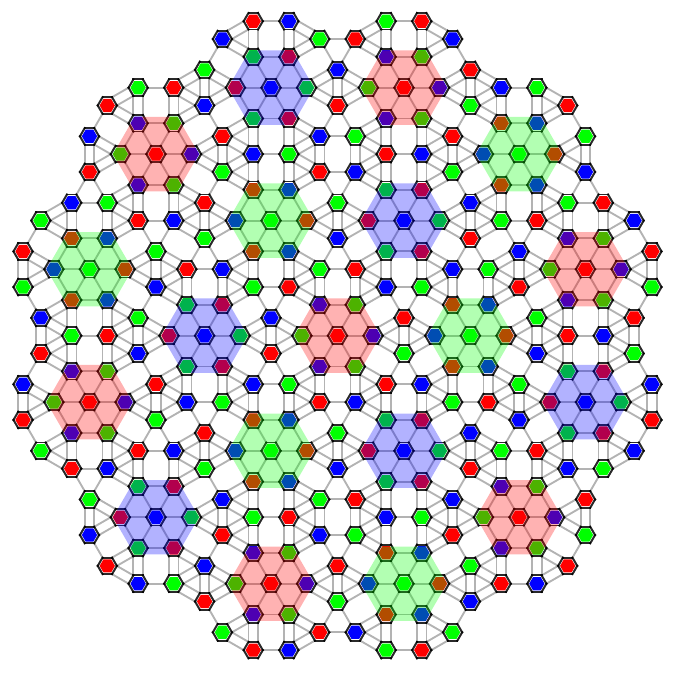} 
\caption{(Left) Portion of the 6-fold QC showing a three-color cluster antiferromagnetic state. Hexagons are colored according to the direction of the variable T. The crosses mark all the bonds corresponding to the perturbation terms $H_b$ defined in Eq.\ref{eq:hamqc2}. (Right) Large colored disks indicate the grouping of hexagons to form block spin variables after one inflation. The color coding of the disks is chosen to reflect the orientations of the block spins (which are aligned antiparallel to the spin of the central hexagon). }
\label{fig:threecolorQC}
\end{figure*}

Putting together solutions for neighboring triangles, one can construct a global 3-color state which minimizes the energy of all the triangles. One such state is shown in Fig.\ref{fig:threecolorQC}, where hexagons have been colored according to their spin direction. This state has a red cluster at the origin. There are two other choices for the color at the center. In addition, for each choice, the state has a mirror equivalent. The simple 3-color state is very close to being a ground state, but it is not an exact ground state of $H^{hex-QC}$ due to the $H_b$ term, which causes some spins to rotate with respect to the ideal three color state. The bonds which correspond to $H_b$ have been marked with a cross in the figure. It can be seen that there are not many such bonds and that they only concern a small number of sites. In addition, for small $r$, the deviations from the 3-color axis of certain spins will be small, and should be negligible away from the region involved. These considerations lead to the conclusion that, in the small $r$ limit, the ground state of the system should approach a 3-color state of the type shown. 

Neglecting the single bond contributions since, as shown in Appendix C, the fraction of these bonds is less than 2 $\%$, we can get a lower bound for the ground state energy for a sample of size $N=6V^{(n-1)}$ sites. Counting up the contributions due to hexagons and overcounting the contributions of the triangles of each kind, we get 
\begin{eqnarray}
    E_0^{hex-QC} (r,+-) &>& -J_1 S^2 (N + \frac{3r}{2} N_{t1}^{(n-1)} + 3r N_{t2}^{(n-1)}) \nonumber \\
    &= & -J_1 S^2 (N + \frac{9}{4} N_{t}^{(n-1)}) \nonumber \\
    &=& - NJ_1 S^2 (1 + \frac{3}{2\lambda})    
    \label{eq:gs5}
\end{eqnarray}
where we used the fact that i) the number of clusters is equal to the number of sites of the tiling of one less generation, $N^{(n-1)}$, then that ii) the number of triangles with single $J_2$ bonds is equal to the number of triangles of type $T_1$, $N_{t1}^{(n)}$, that iii) the number of triangles with double $J_2$ bonds is equal to the number of triangles with single bonds, and iv) the number of vertices in the $n$th patch is equal to $N_{s}^{(n)}+\frac{1}{2}N_{t}^{(n)}$. The correction term involves $N_s$, the number of single bonds. This number is estimated in the infinite size limit (see Appendix C) to be of the order of 2$\%$ of the total number of bonds, and has been therefore neglected.

The long range antiferromagnetic order shown in Fig.\ref{fig:threecolorQC} is expected to be stable for $r \ll 1$. The solution shown in the patch can be continued out to infinite distances. This follows because this state has a self-similar structure under scale change. One can define hexagons on a bigger scale by combining groups of hexagons, as shown by the transparent disks in Fig.\ref{fig:threecolorQC} (right). The new hexagonal blocks are a factor $2+\sqrt{3}$ larger, are composed of 7 clusters and contain 42 spins. It is easy to see that the total (block) spin of the new hexagon is ${\bf{T}}' = -2 {\bf{T}}_0$, where $T_0$ is the spin of the central hexagon. As can be seen in the figure, the colors of the magnetic variables in the larger scale structure are the same as those of the original state, upto a reflection.

\begin{figure*}[h!]
\includegraphics[width=0.45\textwidth]{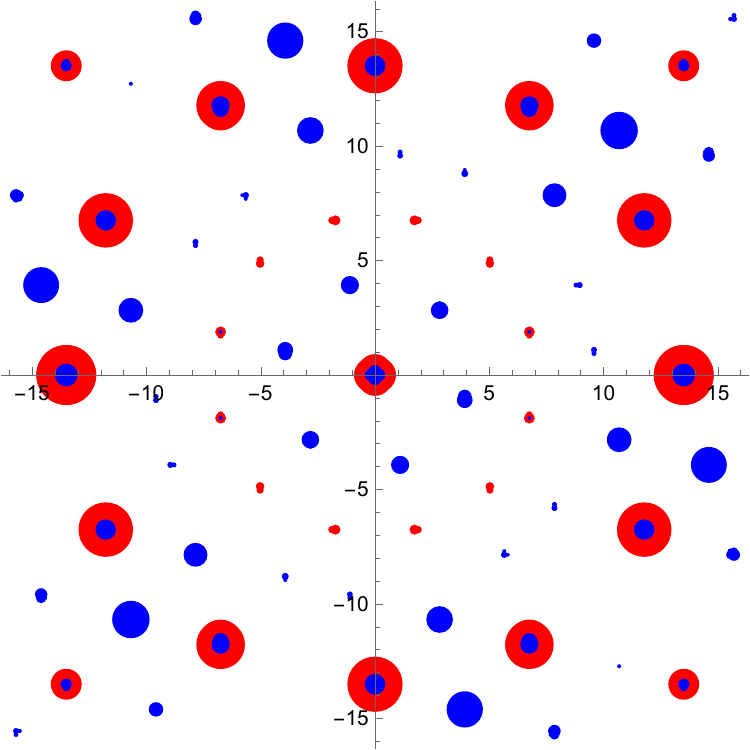}  
\caption{Comparison of structure factors of the nuclear and of the magnetic structure for the finite patch shown in Fig.\ref{fig:threecolorQC}a). The radii of circles are proportional to the nuclear structure factor $S({\bf{Q}}$ (red) and magnetic structure factor $S_{mag}({\bf{Q}})$, computed for a patch of 289 spins in the 3-color state. The $Q=0$ peak is not shown, and the two structure factors have been normalized with different multiplicative factors.}
\label{fig:strucfacQC}
\end{figure*}

In Fig.\ref{fig:strucfacQC}, the  Fourier transform of the spin-spin correlation function in the magnetic state, $S_{mag}({\bf{Q}})$, is compared with the nuclear structure factor, $S({\bf{Q}})$. The radii of the circles, colored red (nuclear) and blue (magnetic), are proportional to the magnitudes of the structure factors (however the scale factors for the two colors are different). The peak at $Q=0$ has been subtracted in both the nuclear and magnetic cases. However, there are peaks very close to the origin, and these overlap around $Q=0$ in the figure. One sees that the (approximate) 12-fold symmetry of the nuclear structure factor is reduced to a 6-fold one, rotated by $\pi/12$, for the magnetic structure factor. These intensities were computed for a patch comprising 289 clusters. The blue spots are rotated by $\pi/12$ with respect to the original axes. This is explained by the tilt of the three sublattices of this magnetic structure. As described in Appendix D, in each one of the three subtilings, the lines of highest density are aligned along lines of slopes $15^\circ, 45^\circ$ and $75^\circ$. 

\subsubsection{Infinite cluster and fluctuating islands mixed phase.}
\begin{figure*}[h!]
\includegraphics[width=0.65\textwidth]{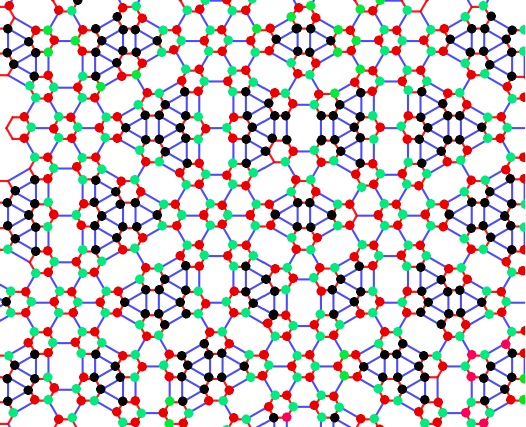} \hskip 1cm
\caption{A region of the 6-fold quasicrystal showing the sites colored according to their nature. Red (sublattice 1) and green sites (sublattice 2) belong to the bipartite infinite network. Enclosed within this infinite network are the islands of black sites which form small finite domains of spins connected by triangles of $J_2$ bonds.  }
\label{fig:smallpatch}
\end{figure*}

Another novel type of long range ordered magnetic phase can occur in the $--$ quadrant. This phase can be described as a collinear antiferromagnet, in which are embedded islands of more weakly ordered spins. This type of state is most easily described in the limit of $\vert J_1\vert \gg \vert J_2\vert$. In the $hex-QC$ model sites occupied by spins fall into two categories. The first category of sites form an infinite, multiply connected, percolating cluster. This set of sites is bipartite, meaning that sites can be divided into sublattices such that the Hamiltonian only couples sites belonging to different sublattices. Therefore, for antiferromagnetic couplings, spins on the infinite percolating cluster can minimize their energy by forming a Néel type state. This is illustrated in the Fig.\ref{fig:smallpatch} (left) which shows the infinite cluster antiferromagnet using red and green to distinguish the two sublattices. 

The remaining sites (colored black in the figure) form small patches of $T_2$ triangles. These frustrated patches are disjoint (do not overlap) being surrounded by sites belonging to the infinite network. It can be shown, with the help of inflation rules for this structure, that there are only three types of patches. They are distinguished by the number of triangles inside the patch, namely i) 2-triangle patch shaped like a diamond ii) 4-triangle patch in the form of a pyramid and iii) 6-triangle patch composed of two groups of 3-triangles forming a butterfly shape. Appendix E provides some more details on the patches.

For antiferromagnetic $J_2$, the triangles lead to frustration whatever the sign of $J_1$. The spins within a patch can order locally, and lie within planes which can vary in a random way from one frustrated patch to another. Examples of spin configurations in some patches are given in Appendix E.  The fraction of frustrated spins compared to the total number of spins is estimated, in Appendix E, to be about 27$\%$. This is fairly large and yet,  the important point is that long range order is nevertheless present in this system, thanks to the fact that these small domains are surrounded by an unfrustrated infinite network. Fig.\ref{fig:smallpatch} shows a finite region of the quasicrystal and its decomposition into the unfrustrated bipartite infinite cluster (vertices decorated by red and green circles) and the frustrated patches that it encloses (vertices decorated by black circles). To understand the ordering, consider first only the spins belonging in the infinite network. At $T=0$, the correlation length should be divergent and long range order present for the infinite cluster. If the frustrated patches were not present, the spins would order in a collinear fashion, parallel to each other if $J_1<0$ or with alternating sign if $J_1>0$.  However, for $r\neq 0$, this strict collinearity is perturbed when the spin is on the boundary of a frustrated patch. 

\begin{figure*}[h!]
\includegraphics[width=0.3\textwidth]{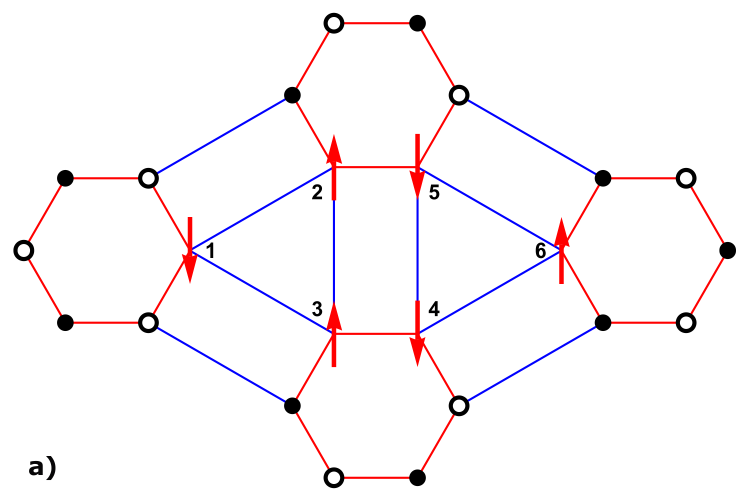} \hskip 0.5cm
\includegraphics[width=0.3\textwidth]{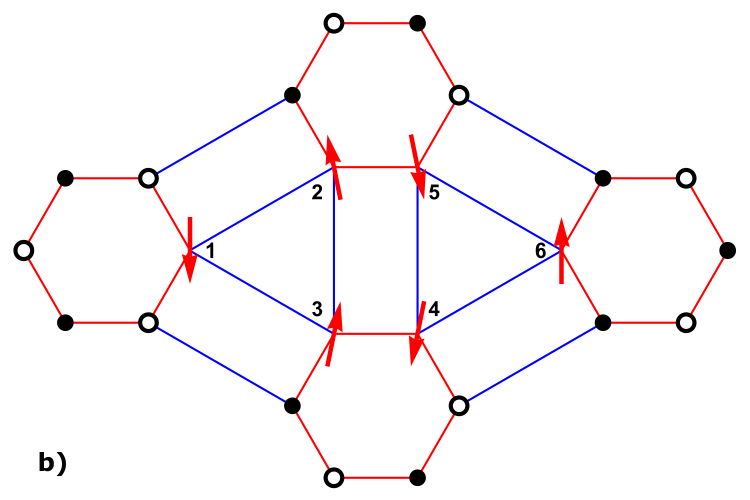} \hskip 0.5cm
\includegraphics[width=0.3\textwidth]{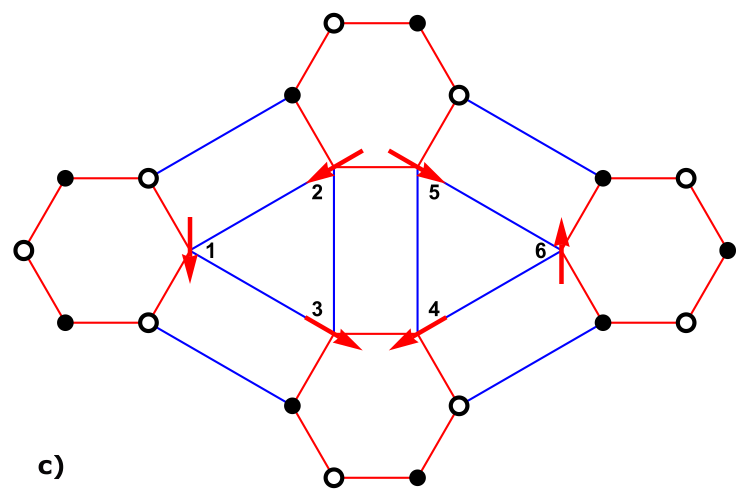} 
\caption{A 2-triangle frustrated patch, showing the interior sites (labeled from 1 through 6) and the exterior sites, whose spins are oriented up (white circles) and down (black circles) in alternation. Spin configurations are shown as function of increasing $r$ in the rigid backbone approximation (see text) a) Configuration for vanishingly small $r$. b) Configuration for a small $r$ value, when the spins are tilted by $\pm\theta$ with respect to their original axis. c) Configuration  for large $r$, when $\theta=\pi/3$. The coplanar ordering is shown to be in the plane of the figure, for convenience. }
\label{fig:twotrimain}
\end{figure*}

As seen in Figs.\ref{fig:twotrimain}, the smallest frustrated patches have two triangles, composed of 6 interior spins, surrounded by the infinite network of collinear alternating up-down spins. Assuming rigid boundary conditions where the boundary spins do not deviate from the spontaneous broken symmetry axis, we show below that the inner spins should have a collinear alignment for small $r$ and go over to a planar arrangement as $r$ increases.

The evolution of the ground state with increasing $r$ is illustrated for the 2-triangle patch in Figs.\ref{fig:twotrimain}. The figures show the patch, which comprises 6 spins in its interior (on the sites labeled from 1 to 6), along with the external spins belonging to the infinite cluster. The collinear up (down) spin directions are shown by white (black) circles respectively. Bonds are colored differently for the $J_1$ (red) and $J_2$ (blue) bonds. For $r=0$, the collinear state, where interior spins have up/down orientations as shown in Fig.\ref{fig:twotrimain}a) minimizes the energy of the patch. For non-zero $r$, and using the symmetry of the patch, one can write the energy of the patch in terms of a single tilt angle $\theta$. For small $r$, the spins tilt, as shown in Fig.\ref{fig:twotrimain}b) by  angles $+\theta$ (for the spins on sites 3 and 4) and $+\theta$ (for the spins on sites 2 and 5). The plane of the tilt angles is arbitrary but, for clarity, the tilts are shown in the plane of the figure. Minimization of the patch energy with respect to $\theta$ yields the tilt angle as a function of $r$. Then, within this ``rigid backbone" approximation, the collinear and the tilted state energies for the 6 spins are
\begin{eqnarray}
   E_{coll}/J_1 &=& -10 -2r \\ \nonumber 
   E_{tilt}/J_1 &=& -8 - 3r - \frac{1}{r} 
\end{eqnarray}
The energies of the collinear and tilted states have a crossing at $r=1$. 
The ground state is given by the collinear solution for  $r\leq 1$, after which the tilted solution becomes more favorable. For $1 \leq r < \infty$, and with the rigid boundary approximation one has
\begin{eqnarray}
   \theta = \arccos(\frac{1+r}{2r}) 
\end{eqnarray}
As $r$ is increased from 1, the tilt angle increases from 0 and reaches the limiting value of $\theta=\pi/3$ , as shown in Fig.\ref{fig:twotrimain}c). This value corresponds to a planar 3 color A-B-C type arrangement of the spins on the first of the triangles, and its time reversed configuration on the second triangle. The plane contains the collinear spin direction but is otherwise arbitrary, leading to disorder.

The spins within patches are largely decorrelated from patch to patch. They can be expected to be only weakly ordered due to the out-of-plane fluctuation modes, similar to the weathervane modes of the previous section. The infinite cluster in contrast, is ordered at $T=0$, and for small $r$, there should persist a high degree of collinearity of spins in the infinite cluster except at the sites immediately adjoining the frustrated islands. There will presumably be a ``healing length" in the deviation from collinearity, as a function of $r$. This length scale remains to be determined by numerical computations. Furthermore, as $r$ is increased, instabilities could occur in favor of a different type of ordering. 

This type of mixed phase -- with a strongly ordered infinite cluster and weakly disordered islands -- is reminiscent of states observed in the Penrose tiling \cite{vedmedenko2003}. For the Penrose tiling, long range dipolar interactions lead to a set of mixed states, with a small fraction of the spins ordering so as to form closed decagonal loops, and the remaining spins remaining disordered, as was checked directly by experiment. The physics underlying those dipolar interaction induced mixed states and our geometrical frustration induced mixed states are however very different.

We conclude this section by stressing that the two solutions given above are approximate solutions. They should be close to the ground states for small $r$, and possibly can be extended out to a larger region of the phase diagram. This would need to be clarified by numerical computations. In the remainder of the phase space, more complex forms of order, as well as disordered phases are likely to exist. For antiferromagnetic $J_2<2J_1$, the local constraints on individual triangles Eq.\ref{eq:constraint} can always be satisfied but the nontrivial problem consists of minimizing all triangles. It will be interesting to study whether such a solution exists, possibly non-coplanar, and possibly a spin liquid, as in the case of distorted Kagome models \cite{viswanath, reuther} or the disordered Kagome model \cite{moessner}.  

\section{Discussion and perspectives}
The examples considered in this paper have illustrated various types of long range magnetic order -- commensurate FM and AFM phases and incommensurate phases in periodic structures which are closely related to quasicrystals. All our examples are based on square triangle tilings, and have spins that lie on vertices of hexagons. The periodic systems have  a variety of local bond configurations, which resemble local configurations found in the quasicrystal, but they of course differ from the latter at larger distances. We have shown that frustration can give rise to complex phase diagrams in the periodic systems. 

More interestingly, in the case of the 6-fold quasicrystal, we have given two examples of magnetic phases with quasiperiodic long range order. These magnetic states are the first nontrivial ones described in a frustrated quasiperiodic system. The first state is a three-color magnetic state, where spin clusters are aligned along one of three directions, and each color forms a sublattice rotated at $\pi/12$ with respect to the nonmagnetic structure. The second magnetic phase, formed when inter-cluster couplings are antiferromagnetic, is a mixed state in which strong and weak magnetic ordering co-exist. The strongly correlated spins form percolating cluster which surrounds small regions of more weakly correlated spins. It will be interesting to study response of this state under applied fields, and the nature of magnetic excitations.

We have shown that the family of magnetic states comprises collinear, coplanar and three-dimensional ground state spin configurations. Among the collinear states alone there are different kinds of ferromagnetic and antiferromagnetic states. Each one can be described by a suitably defined order parameter and could be detected by their differing responses to external magnetic fields for each of these systems. The study of low energy excitations and thermodynamic properties is an interesting direction for future work, with relevance to experiments.

Another direction for study concerns the dependence of spin-spin interactions on distance. While the standard RKKY formula may hold at short distances, especially in small approximants, it should be pointed out that, especially in the quasicrystal, the standard RKKY interactions could be significantly modified due to the aperiodicity of the potential. The modified form of interactions could be computed using perturbation theory, probably with results somewhat analogous to the ones found for RKKY interactions in weakly disordered metals \cite{rkky1}. Non-Heisenberg models and the role of anisotropic interactions are important extensions of the study in order to understand other systems based on rare earth ions such as Tb.

The 6-fold quasicrystal has a better known cousin, the dodecagonal quasicrystal, which has been invoked to describe a number of experimental systems. This latter has local environments which resemble those seen in periodic lattices, however, they are combined to form complex patterns. The resulting bond connectivity and frustration will pose a challenging problem to study in the future.

In addition, a number of experimentally relevant extensions of the models remain to be addressed. One of the points to address is the fact that we have considered here only clusters which are not intrinsically frustrated. Our focus has been to study the effects of frustration induced by the long range connectivity and differences between periodic and quasiperiodic structures for clusters which are maximally ordered in the decoupled limit. However, in the next step, it will be important to introduce next nearest neighbor intra-cluster couplings. These lead to frustration within each of the clusters, which in turn affect long range order. The effects of adding second and third nearest neighbor interactions are well-known for periodic lattices such as honeycomb and Kagome lattices \cite{kallin,messio} where they can lead to many phases, including out-of-plane ``tetrahedral" or ``umbrella" states. Our systems bear similarities to the square-Kagome or shuriken antiferromagnets, whose magnetic properties have been reported recently \cite{gembe}. These propreties of extensive degeneracy, order-by-disorder and non-coplanar ordering are expected to also occur in our frustrated models. 

One of the open questions concerns spin glass transitions, and non-equilibrium low temperature phases. This type of phase has been observed experimentally in those periodic approximants which have negative $T_{cw}$, that is, predominantly antiferromagnetic interactions. Quasicrystals in general tend to show spin glass ordering, with some notable exceptions that we have already mentioned. Frustration clearly plays a role, but it is also an interesting question as to the role of disorder in the formation of spin glass phase in these alloy systems.

Quantum effects have been neglected in view of the large local moments of the Gd ions in the quasiperiodic alloys of interest here. These would however become pertinent for small values of the spin, raising questions concerning the stability of ordered states with respect to quantum fluctuations. One standard theoretical approach, spin wave theory, can be used for many of the ordered periodic ground states that we have listed, along the lines of what was done in \cite{jagamoess}. It should also be possible to carry out a spin wave expansion for the 3-color quasiperiodic ground states, using an approach similar to that used for the Ammann-Beenker \cite{milat} and the Penrose \cite{szallas2} tilings. The disordered phases, less amenable to analytical or numerical techniques are of course also the most interesting from the point of view of spin liquid formation. 

To finish, the goal of these studies has been to get an understanding of the magnetic ordering in 3D quasicrystals. The experimental systems, where spins are majoritarily located on icosahedral clusters, are frustrated due to both their geometry as well as due to competition between long range RKKY couplings. Our simplified 2D models were conceived to mimic these properties, and they shed light on the way that magnetic structures change, depending on the bond configurations at short range. As the unit cell size increases, the states are more complex reflecting the connectivity at longer ranges. In the quasiperiodic limit, this study shows that there could be long range magnetic order in certain limits, such as when intra-cluster interactions dominate and give rise to cluster-antiferromagnets or ferromagnets. More generally, with fine tuning of interactions there may result other novel types of quasiperiodically ordered magnets. These questions remain for future investigations.

\vskip 0.5cm
{\bf{Acknowledgments}} I would like to thank R. Tamura, F. Labib and T. Sugimoto for many useful discussions and for kindly giving me access to their unpublished experimental data.

\vskip 0.5cm

\appendix
\section{Experimental phase diagram for a family of magnetic approximants}
\begin{figure*}[h!]
\includegraphics[width=0.6\textwidth]{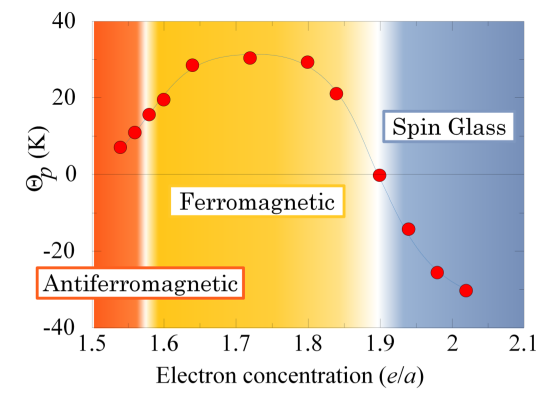} \hskip 1cm 
\caption{An experimental phase diagram obtained for the family of approximant crystals, Au$_x$Al$_{100-x}$Gd$_{14}$ showing the evolution of the Curie-Weiss temperature as a function of the number of electrons per atom $e/a$. The three colored regions indicate the nature of the low temperature magnetic ordering in this alloy system. (Figure reproduced with permission from \cite{ternarytamura}.)}
\label{fig:exptl}
\end{figure*}

This appendix provides a brief summary of the experimentally determined magnetic phase diagram of the ternary alloy system Au$_x$Al$_{86-x}$Gd$_{14}$ as reported in \cite{ternarytamura}. Changing $x$ results in a change of electron concentration $e/a$ (electron per atom), within the range of $1.5 < e/a < 2.1$. The high temperature magnetic susceptibilitties were fitted to find the Curie-Weiss temperatures, $T_{CW}$. Due to the long range RKKY couplings, the Curie-Weiss temperature evolves in a smooth  fashion as compositions $x$ was varied. Fig.\ref{fig:exptl} shows a plot of $T_{CW}$ versus $e/a$ for a set of samples. The colors show the nature of the low temperature ordering. 
It can be seen that the samples are initially antiferromagnets for the smallest $e/a$, then become ferromagnets, and finally for the largest $e/a$, where the Curie-Weiss temperature is large and negative, are spin glasses showing non-equilibrium behaviors. 

The above phase diagram indicates that the Curie-Weiss temperature is an indicator of the nature of low temperature magnetic order. Interestingly, a large number of related (isostructural) compounds have been found, subsequently, to fit the same phase diagram. Larger approximants, when they exist, also fit the pattern seen in Fig.\ref{fig:exptl} as shown in \cite{tamura2appxt}. The quasicrystalline antiferromagnet reported in \cite{tamura2025}, however, does not.

\section{Statistics of tiles in the 6-fold quasicrystal}
\begin{figure*}[h!]
\includegraphics[width=0.2\textwidth]{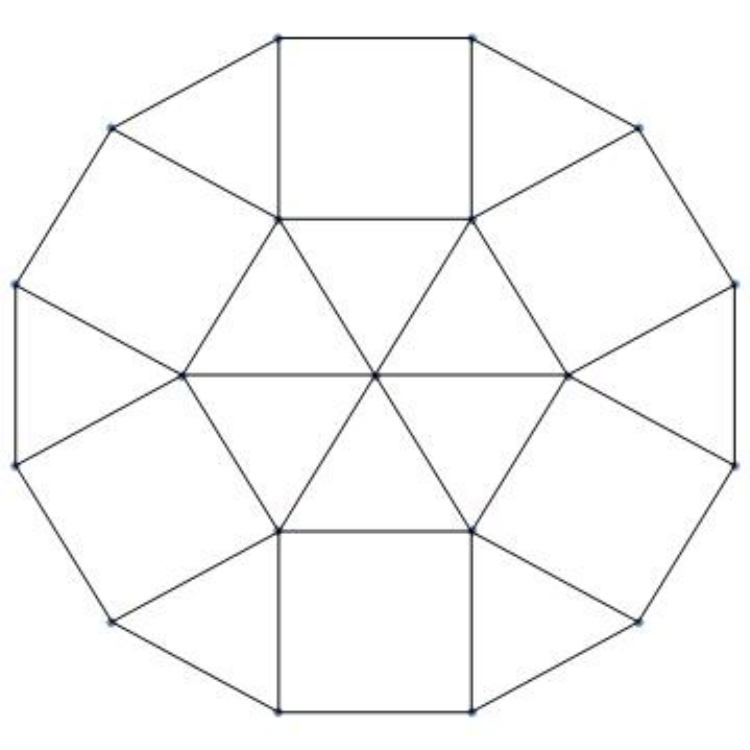} \hskip 1cm \includegraphics[width=0.5\textwidth]{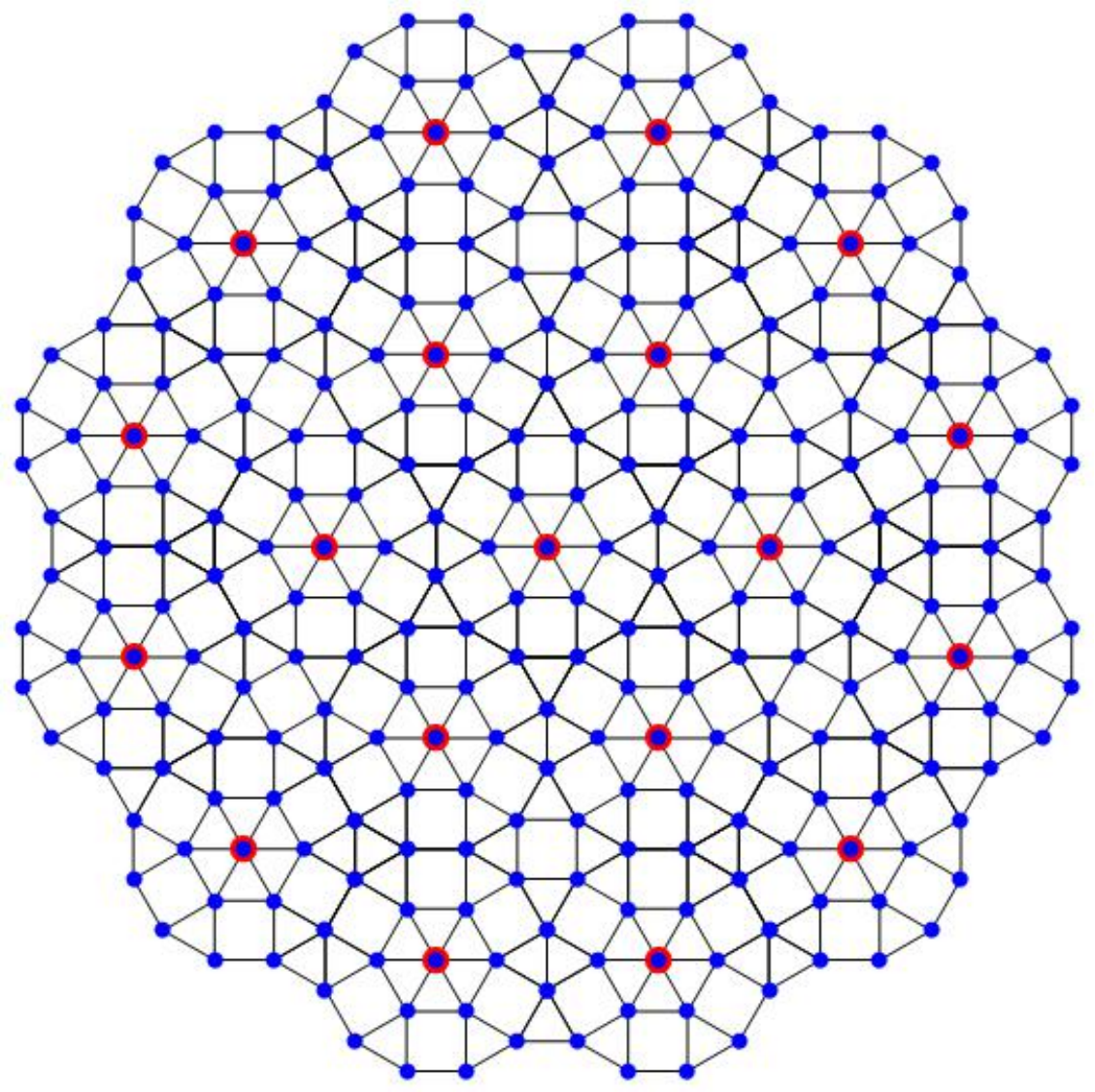}
\caption{(Left) A decagonal seed. (Right) The patch obtained after one inflation (see text).}
\label{fig:inflation}
\end{figure*}

As mentioned in the main text, the tiling can be obtained by a method of substitution. The initial seed is a dodecagon formed of 12 triangles and 6 squares, as shown in the left hand figure of Fig.\ref{fig:inflation}. One can assume, for convenience, that the edge lengths are unity. In step 1, all the edges of this dodecagonal seed are expanded by a linear scale factor $\lambda=2+\sqrt{3}$ -- these points are indicated by red circles in the center figure of Fig.\ref{fig:inflation}. Next, each of the vertices of this bigger tiling is decorated with a seed dodecagon. Doing so results in a bigger square triangle patch with no overlaps or holes, as in the center figure of Fig.\ref{fig:inflation}. We term this operation an ``inflation" since the patch gets larger (however one may find in the literature the term deflation used for this same operation). One obtains the quasicrystal covering the entire 2D plane in the limit of infinite inflations. 
After a finite number of inflations, the number of squares and triangles in the bulk of the tiling (excluding boundary effects) can be counted as follows: let the number of squares and triangles in the $n$th patch be denoted by $N^{(n)}_s$ and $N^{(n)}_t$ respectively (with $N^{(0)}_s$=6 and $N^{(0)}_t=12$). The total area for the $n$th patch can be written in terms of these numbers, and related to the area of the preceding patch as follows:
\begin{eqnarray}
    A^{(n)}& =& N^{(n)}_s + \frac{\sqrt{3}}{4} N^{(n)}_t \nonumber \\
    &=& (2+\sqrt{3})^2 A^{(n-1)} \nonumber \\
    &=& (2+\sqrt{3})^2 (N^{(n-1)}_s + \frac{\sqrt{3}}{4} N^{(n-1)}_t) \nonumber \\
    &=& (7 N^{(n-1)}_s + 3 N^{(n-1)}_t) + \frac{\sqrt{3}}{4} (16 N^{(n-1)}_s + 7 N^{(n-1)}_t)
\end{eqnarray}
where the second line follows from the length scale of inflation. The last relation shows that the number of squares and of triangles before and after inflation are related as 
\begin{eqnarray}
   \left( \begin{array}{c}
       N^{(n)}_s    \\
         N^{(n)}_t
    \end{array}  \right)
    = 
   \left( \begin{array}{cc}
    7 & 3 \\
    16 & 7
\end{array} \right)
\left(\begin{array}{c}
       N^{(n-1)}_s    \\
         N^{(n-1)}_t
    \end{array}  \right)
\end{eqnarray}
Solving this system gives the Perron-Frobenius eigenvalue, $\lambda^2=(7+4\sqrt{3})$, which describes the growth of the number of tiles in successive patches. The associated eigenvector which gives the ratio of number of squares to that of triangles, $\sqrt{3}/4$, in the limit $n\rightarrow \infty$. There are two types of triangles, type 1 having a horizontal edge and type 2 having a vertical edge. They occur in equal proportions.

Finally, using the Euler relation, $V=2+E-F$ which related the number of vertices (V) to the number of edges (E) and faces (F), one can write the total number of sites in a large patch of the $n\gg 1$ th generation as follows:
\begin{eqnarray}
    V^{(n-1)} = 2 + N_s^{(n-1)} + \frac{1}{2} N_t^{(n-1)}
\end{eqnarray}

\section{Fraction of single bonds in the Hamiltonian $H^{hex-QC}$}
In Eq.\ref{eq:hamqc2} the effective Hamiltonian was separated into a term involving triangles, and a term consisting of the small number of single bonds not belonging to any triangle. Let the hexagonal clusters be defined on a patch of tiling of generation $n$. The $n=2$ case is shown in Fig.\ref{fig:threecolorQC} with the single bonds marked by an "X". These are the bonds shared by two squares placed side by side. We show here how to get an estimate of the number of such single bonds as a fraction of the total number of bonds. 

\begin{figure*}[h!]
\includegraphics[width=0.6\textwidth]{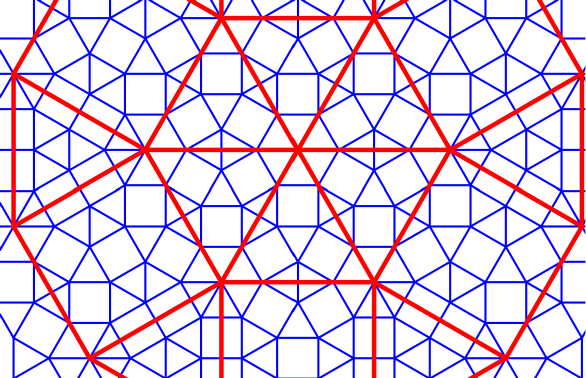} 
\caption{Close-up of tilings before and after inflation, showing the transformation of squares and triangles into larger patches of squares and triangles respectively. Note the different transformations in the case of the $T_1$ and $T_2$ triangles.}
\label{fig:rules}
\end{figure*}

The total number of bonds for the $n$th tiling can be estimated to be $\mathcal{N}^{(n)} \sim 2N_s^{(n-1)} + \frac{3}{2} N_{t1}^{(n-1)} + (3+\frac{3}{2})N_{t2}^{(n-1)} = 2N_s^{(n-1)} + \frac{9}{4} N_{t}^{(n-1)}$.
since squares have an average of 2 bonds each, type 1 triangles have $3/2$ bonds, and type 2 triangles have an additional 3 bonds.

To estimate the number of single bonds $\mathcal{N}_b^{(n)}$, we rely on the transformation of tiles under an inflation operation. The transformation of a (square)triangle into a (square)triangle shaped patch of tiles is shown in Fig.\ref{fig:rules}. It can be seen that double-square configurations occur after inflating a triangle of type 2 (having a vertical edge). That is, the number of single bonds in the $n$th generation tiling is related to the number of $T_2$ triangles in the $n-2$th generation

\begin{eqnarray}
    N_b^{(n)} \sim \frac{3}{2}N_{t2}^{(n-2)} = \frac{3}{4}N_{t}^{(n-2)}
\end{eqnarray}
 In the infinite size limit, the fraction of single bonds is therefore
\begin{eqnarray}
    N_b^{(n)}/\mathcal{N}^{(n)} \sim \frac{\frac{3}{4} N_{t}^{(n-2)}/N_{t}^{(n-1)}}{\frac{9}{4}+2 N_s^{(n-1)}/N_t^{(n-1)}} 
    \approx 0.017
\end{eqnarray}
where we used the results that for large $n$, $N_{t}^{(n-1)}/N_{t}^{(n)}$ tends to $1/\lambda^2$ and $N_s^{(n)}/N_t^{(n)}$ tends to $\sqrt{3}/4$.

\section{The 3-color state subtilings structure}
\begin{figure*}[h!]
\includegraphics[width=0.3\textwidth]{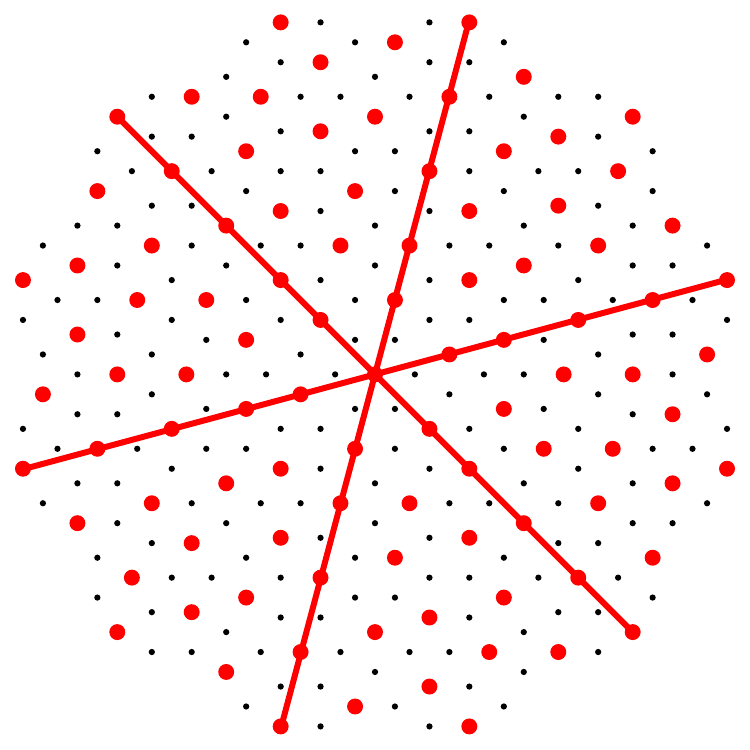}  
\includegraphics[width=0.3\textwidth]{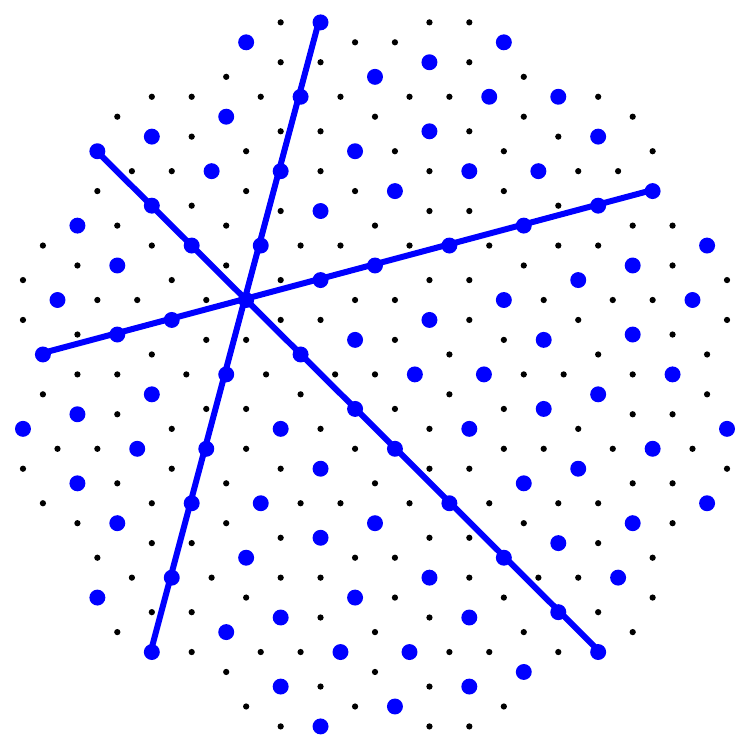}  
\includegraphics[width=0.3\textwidth]{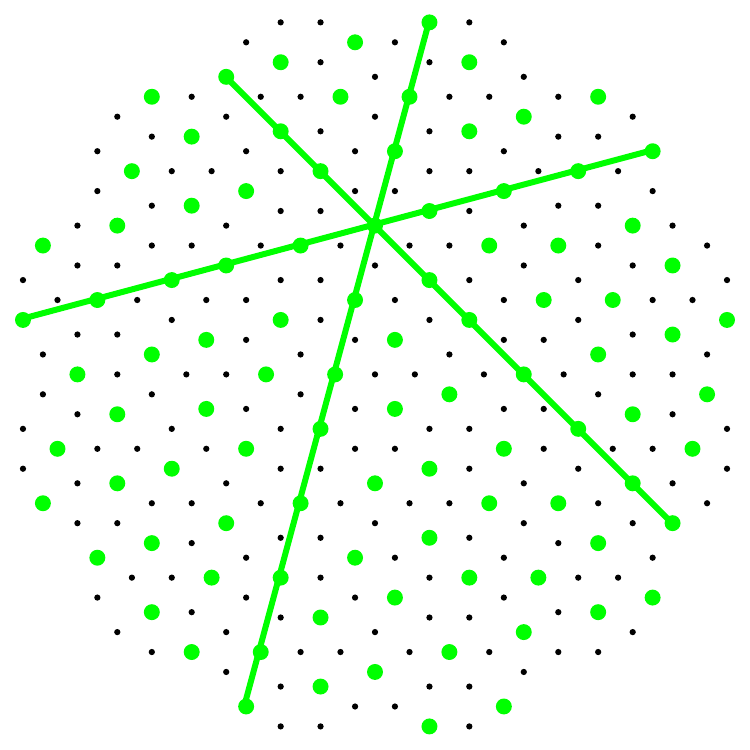}
\caption{Break-up of the 3-color magnetic state shown in Figs.7 highlighting the clusters of each individual sub-tiling. Figures show the red, blue and green cluster positions for a 289 cluster tiling. The lines oriented at angles of $\pi/12,3\pi/12$ and $5\pi/12$ indicate the directions of high density of clusters. They are tilted by $\pi/12$ with respect to the principal directions of the original structure.}
\label{fig:subtilings}
\end{figure*}

The three subtilings formed by red, blue and green clusters of Fig.7 are individually shown in Figs.\ref{fig:subtilings}. It can be seen that points lie on lines of high density which are inclined at angles of $15^\circ, 45^\circ$ and $75^\circ$ with respect to the $x$ axis of the original structure. Due to our choice of ground state, the red sub-tiling has a center of symmetry at the origin, through which the three lines can be seen to cross. For each of the remaining two sublattices, we show points where only two of the lines of maximum density cross. 

The structure factor shown in the main text is rotated by $15^\circ$ and also has lower symmetry (3-fold rather than 6-fold).

\begin{figure*}[h!]
\includegraphics[width=0.5\textwidth]{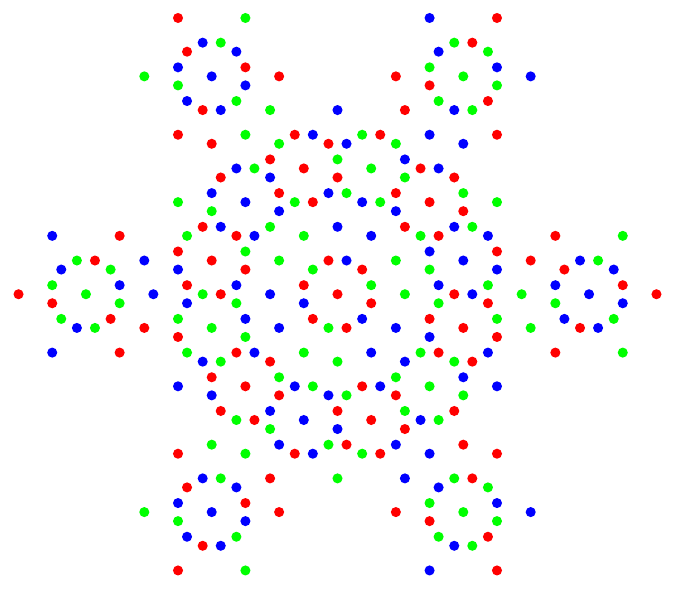} 
\caption{ The 3-color magnetic state of Fig.\ref{fig:threecolorQC} (left) represented in perpendicular space. The points are projections of centers of clusters, their colors correspond to the spin orientation of the cluster.}
\label{fig:perpspace}
\end{figure*}

Finally, it is interesting to ask what the magnetic 3-color state looks like, in an alternative representation of the tiling in so-called perpendicular space. It is possible to lift the vertices of this tiling into four dimensional space as described in \cite{forster2024}, and thence projected onto a perpendicular space. The projections of points of the tiling form a fractal in the perpendicular space, analogous to the 12-fold quasiperiodic tiling, as discussed in \cite{fractalAD1}. The perpendicular space representation of vertices of a colored tiling of 289 sites is shown in Fig.\ref{fig:inflation}(right). Colors correspond to the red-blue-green state shown in Fig.\ref{fig:perpspace}.

\section{Frustrated islands in the mixed phase.}

Fig.\ref{fig:rules} shows a patch of the $n$th tiling after multiplication by $\lambda=2+\sqrt{3}$ in red, with superposed a $n+1$th tiling in blue. It illustrates how squares and triangles of the $n$th generation patch can be decomposed into squares and triangles of the $n+1$th patch. In particular, one sees clusters of $T_2$ triangles which appear with each step of inflation. These are of four types as mentioned in the main text.

As the figures in the main text, Figs.\ref{fig:twotrimain} show, the smallest frustrated patches have two triangles, composed of 6 interior spins, surrounded by the infinite network of collinear alternating up-down spins. As discussed in the main text, the spin configurations inside each patch is rather weakly coupled to those in the neighboring patches. If the backbone is assumed to be perfectly rigid, then the configuration inside the patches can be computed as a function of $r$. For $r < 1$, the spins in the interior of the patch are collinear along the axis of the infinite backbone, as we have already seen in the main text. For $r>1$, the spins in the interior of each patch become tilted, with the tilt angle gradually increasing with $r$. For large $r$, one obtains a planar 3 color state within one triangle, and its time reversed configuration on the second triangle. 

\begin{figure*}[h!]
\includegraphics[width=0.45\textwidth]{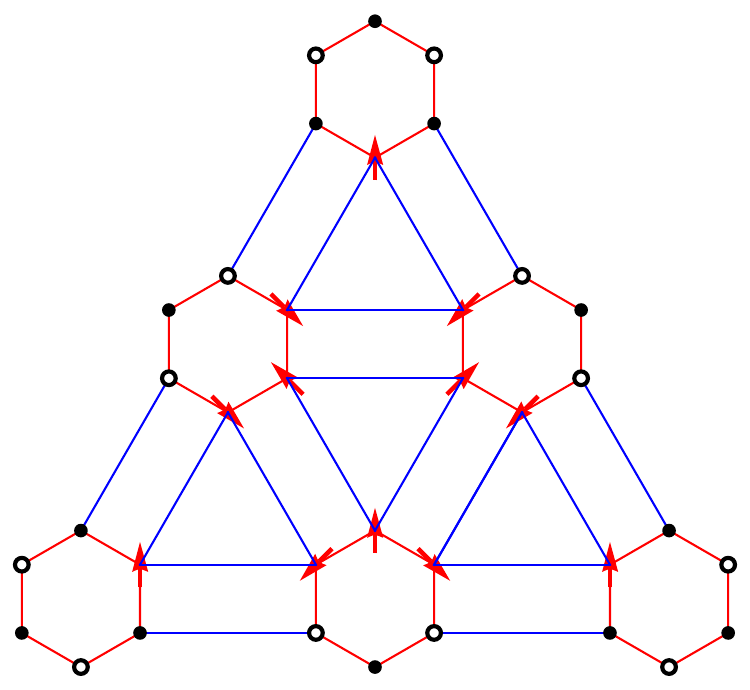} \hskip 0.5cm
\caption{ The second category of frustrated patch of spins has four triangles in its interior. Exterior spins belonging to the backbone are shown with circles colored black (spin up) and white (spin down). Possible spin tiltings for small $r$ values are indicated. }
\label{fig:fourtri}
\end{figure*}

The next smallest frustrated patches have four triangles as shown in Fig.\ref{fig:fourtri}. They are composed of 12 interior spins (in red), surrounded by the infinite network of collinear alternating up-down spins (represented by black and white circles). The arrows indicate a possible minimum energy state of the 12 interior spins for large $r$ assuming rigid collinear exterior spins.

\begin{figure*}[h!]
\includegraphics[width=0.4\textwidth]{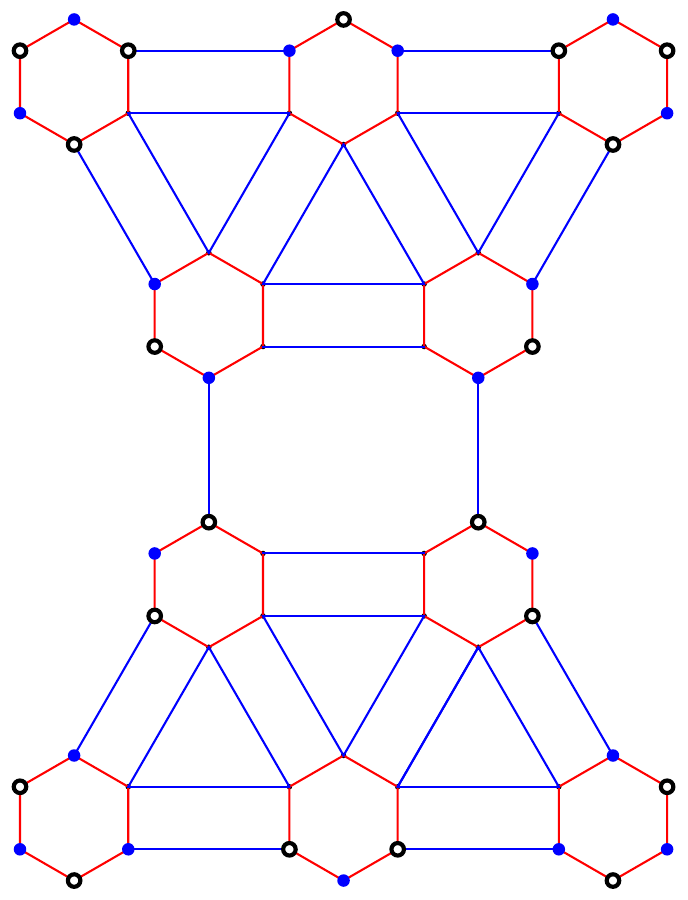} \hskip 0.5cm
\caption{The largest frustrated patch has six triangles, arranged in two groups of three. Exterior spins belonging to the backbone are shown with circles colored black (spin up) and white (spin down). Interior spins directions are not indicated. }
\label{fig:sixtri}
\end{figure*}

The remaining frustrated islands have a symmetric form of 3+3 triangles arranged in a bow-tie pattern as shown in Fig.\ref{fig:sixtri}. In the Néel state described in the text for $J_1<0$, these islands have alternating spins on the boundary and an interior frustrated region of  22 interior spins. 

The inflation rules for squares and triangles, shown in Fig.\ref{fig:rules}, ensure that no other larger frustrated islands occur on the tiling. The ratio of frustrated sites to the total number of sites, $x_f$, can be estimated as follows. Using the Euler relation, $V=2+E-F$ which related the number of vertices (V) to the number of edges (E) and faces (F), one can write the total number of sites in a large patch of the $n\ll 1$th generation as follows
\begin{eqnarray}
   N = 6 V^{(n-1)} \approx 6 (N_s^{(n-1)} + \frac{1}{2} N_t^{(n-1)})
\end{eqnarray}
using the fact that spins lie on hexagons of the tiling, which in turn map to vertices of the $n-1$th tiling.
 As for the frustrated sites, they are related to the number of $T_2$ triangles by $\sim \frac{3}{2}N_t^{(n-1)}$. Therefore the fraction of frustrated sites $x_f$ is
\begin{eqnarray}
    x_f = \frac{(3/2)N_t^{(n-1)}}{6(N_s^{(n-1)}+ (1/2)N_t^{(n-1)})} = \frac{1}{2+\sqrt{3}}
\end{eqnarray}
The total fraction of the frustrated islands in the infinite tiling is therefore $x_f = \lambda^{-1} \approx 0.27$.

\end{document}